\documentclass[a4paper,fleqn]{cas-dc}
\usepackage{enumitem}
\usepackage{algorithm}
\usepackage{algorithmic}
\usepackage{float}
\usepackage{caption}
\usepackage{placeins}
\usepackage{cuted}
\usepackage{graphicx}
\usepackage[T1]{fontenc}
\usepackage[utf8]{inputenc}
\usepackage{adjustbox}
\usepackage{booktabs}
\usepackage{tabularx}
\usepackage{array}

\usepackage[numbers]{natbib}
\def\tsc#1{\csdef{#1}{\textsc{\lowercase{#1}}\xspace}}
\tsc{WGM}
\tsc{QE}
\begin{document}
\raggedbottom
\let\WriteBookmarks\relax
\def\floatpagepagefraction{1}
\def\textpagefraction{.001}

\shorttitle{DDQN-MLP for Robust Ransomware Detection}

\shortauthors{Ferdous et al.}

\title[mode=title]{DDQN-MLP: An Explainable and Adversarially Robust DRL-Guided Adaptive Learning Framework for Ransomware Detection}

\author[1]{Jannatul Ferdous}[orcid=0000-0002-9612-0482]
\cormark[1]
\ead{jferdous@csu.edu.au}
\credit{Conceptualization, Methodology, Software, Investigation, Writing -- Original Draft}

\affiliation[1]{organization={School of Computing, Mathematics and Engineering, Charles Sturt University},
            city={Wagga Wagga},
            postcode={2650},
            state={NSW},
            country={Australia}}

\author[2]{Rafiqul Islam}
\credit{Supervision, Validation, Writing -- Review \& Editing}

\affiliation[2]{organization={School of Computing, Mathematics and Engineering, Charles Sturt University},
            city={Albury},
            postcode={2640},
            state={NSW},
            country={Australia}}

\author[3]{Arash Mahboubi}
\credit{Validation, Writing -- Review \& Editing}

\affiliation[3]{organization={School of Professional Studies, University of New South Wales},
            city={Canberra},
            state={ACT},
            country={Australia}}

\author[4,5]{Md Zahidul Islam}
\credit{Supervision, Writing -- Review \& Editing}

\affiliation[4]{organization={School of Computing, Mathematics and Engineering, Charles Sturt University},
            city={Bathurst},
            postcode={2795},
            state={NSW},
            country={Australia}}

\affiliation[5]{organization={AI and Cyber Futures Center, Charles Sturt University},
            addressline={Panorama Avenue},
            city={Bathurst},
            postcode={2795},
            state={NSW},
            country={Australia}}

\cortext[1]{Corresponding author}

\begin{abstract}
Ransomware detection remains a critical challenge because modern variants often exhibit diverse, evasive, and partially benign-like behavioral patterns. Conventional machine learning approaches rely on static supervised training objectives, where all samples are treated uniformly during optimization. These fixed strategies often fail to adapt to latent behavioral heterogeneity and struggle when subtle ransomware behaviors overlap significantly with benign activity. To address this limitation, this study introduces an intelligent training-time Deep Reinforcement Learning (DRL) framework for behavioral ransomware detection using modern Windows 11 sandbox telemetry. The proposed system employs a Double Deep Q-Network (DDQN) as an adaptive, discrete sample-weighting controller that models the classifier optimization sequence as a Markov Decision Process. By observing batch-level loss trajectories and prediction confidence dynamics, the DDQN agent dynamically optimizes the sample-importance weights to guide a lightweight Multilayer Perceptron (MLP) backbone. Crucially, after training, the DDQN component is fully decoupled, ensuring that only the lean, resource-efficient MLP is deployed for low-latency production inference. Evaluated via 5-fold stratified cross-validation on a balanced dataset of 2,000 executable profiles spanning 30 ransomware families, the DDQN-MLP framework demonstrates superior and stable optimization performance, achieving an accuracy of 99.30\%, an F1-score of 0.9930, and an ROC-AUC of 0.9991 over conventional static, focal-loss, and alternative DRL variants. Post-hoc model interpretability was verified using SHAP and LIME interpretability methods, augmented by a novel SHAP-gradient alignment diagnostic to guarantee explanation consistency with model sensitivity. Furthermore, empirical white-box adversarial stress testing across multiple perturbation magnitudes confirms that integrated adversarial training successfully reinforces feature space robustness without degrading clean data accuracy. The findings demonstrate that DDQN-MLP provides a robust, explainable, and computationally efficient expert system for high-throughput ransomware defense.

\end{abstract}


%\nocite{*}

% Keywords

\maketitle

% Main text

\section{Introduction}

Effective ransomware detection necessitates not only the selection of an appropriate classifier architecture but also the capacity to learn robustly from behaviorally heterogeneous datasets. Ransomware has become one of the most disruptive and financially damaging threats to modern computing infrastructure~\cite{ref1}; however, its complexity extends beyond its raw prevalence. In contrast to conventional malware, which primarily targets data theft or corruption, ransomware employs cryptovirology-based extortion by encrypting files and demanding payment for decryption keys. This approach is further complicated by its ability to mimic legitimate high input/output (I/O) operations, creating a behavioral duality that results in significant operational downtime and imposes unique challenges on supervised learning pipelines. From an optimization perspective, this behavioral mimicry introduces severe class overlap and non-convex boundaries into the feature space. When subtle ransomware traces appear identical to high-I/O benign processes, standard empirical risk minimization causes the classifier's weight updates to be dominated by easily classifiable ``clear-cut'' samples, frequently causing the network to underperform on complex, deceptive boundary cases.

The frequency and sophistication of such attacks have escalated dramatically, with a reported recovery cost of \$1.53 million in 2025~\cite{ref2}, a trajectory amplified by ransomware-as-a-service (RaaS) platforms that democratize attack capabilities and enable even low-skilled threat actors to launch enterprise-grade campaigns~\cite{ref3}. Modern variants such as WannaCry, Ryuk, and LockBit~3.0 further compound this challenge through polymorphism, multistage execution, API misuse, intermittent encryption, and deliberate evasion strategies~\cite{ref4,ref5}. These converging factors, including escalating prevalence, behavioral evasiveness, and the difficulty of learning from heterogeneous signals, highlight the need for detection systems that are not only precise but also adaptively trained to manage the full complexity of contemporary ransomware behavior.

Despite advancements in attack sophistication, most ransomware detection research relies on behavioral datasets collected from outdated Windows environments (Windows~7/8/10) and low-interaction sandboxes such as Cuckoo~\cite{ref6,ref7,ref8,ref9}. These datasets fail to capture Windows~11-specific features, including kernel mitigations, revised registry semantics, modern CryptoAPI behavior, and cloud-integrated security telemetry, resulting in incomplete behavioral traces. Moreover, behavioral ransomware datasets are inherently heterogeneous; some families generate subtle traces resembling benign activity, whereas others exhibit strong malicious signatures. Although traditional machine learning~\cite{ref7,ref10}, deep learning~\cite{ref4,ref9,ref11}, and transfer-learning-based approaches~\cite{ref12,ref13} show promise in ransomware classification, they treat all samples equally during gradient updates. Consequently, in the presence of severe behavioral heterogeneity, conventional training loops suffer from unstable optimization trajectories, gradient variance, and reduced cross-validation stability.

Deep reinforcement learning (DRL) offers a compelling solution to these optimization limitations by modeling the training progression as a sequential control task~\cite{ref14}. Unlike traditional supervised learning, which commits to a rigid, static optimization objective, a DRL control agent can dynamically adjust the importance of individual training instances based on feedback from the classifier's current internal state. In this context, the training trajectory of a neural classifier is formalized as a finite Markov Decision Process (MDP), where the state space maps changing optimization metrics such as batch-level loss and prediction confidence, the actions denote discrete sample-weight alterations, and the reward tracks downstream directional improvements in classification error. Although various DRL paradigms, including value-based networks (DQN, DDQN), policy-gradient methods (PPO), and actor-critic frameworks (A2C), have historically excelled in robotics~\cite{ref15,ref16}, autonomous vehicles~\cite{ref17,ref18}, video games~\cite{ref19,ref20}, and autonomous cyber defense~\cite{ref21,ref22,ref23,ref24}, their application as a dedicated training-time optimization engine for behavioral ransomware detection remains largely unexplored.

To systematically advance this domain, this study investigates and addresses four critical research gaps that currently limit the efficacy and trustworthiness of automated ransomware triage systems.

\begin{enumerate}
\item \textbf{Limited Exploration of DDQN-Based Ransomware Detection:} Although DQN-based models have demonstrated strong performance in intrusion detection~\cite{ref25}, IoT security~\cite{ref26}, botnet detection~\cite{ref27}, phishing detection~\cite{ref28}, and malware analysis~\cite{ref29}, their application to ransomware remains limited. This gap is particularly important because ransomware exhibits distinctive behavioral patterns, including mass encryption, registry manipulation, file locking, and privilege escalation, which are naturally compatible with an MDP formulation.

\item \textbf{Outdated Behavioral Datasets:} Most existing ransomware datasets rely on Cuckoo Sandbox executed on legacy Windows versions (Windows~7/8/10), which fail to capture modern operating-system semantics and restrict real network interactions~\cite{ref8,ref9,ref30,ref31}. Windows~11 introduces modified privilege models, updated registry semantics, enhanced filesystem protections, and new kernel behaviors that legacy environments cannot accurately represent~\cite{ref32}. Moreover, Cuckoo's network isolation limits command-and-control (C2) visibility, resulting in incomplete behavioral traces.

\item \textbf{The Post-Hoc Explanation--Sensitivity Alignment Gap:} Post-hoc explainability techniques such as SHAP and LIME have been explored in prior security analytics and RL-based defense systems~\cite{ref33,ref34}. However, existing systems lack a unified mechanism to verify whether explanation outputs genuinely align with the empirical gradients driving the underlying neural model, creating a critical explanation--optimization disconnect.

\item \textbf{Insufficient Adversarial Robustness and Integrated Trustworthiness:} Although a limited number of DRL-based threat detection studies consider adversarial settings~\cite{ref35,ref36}, the robustness of DRL-driven ransomware detectors against adversarial attacks remains largely unexamined. Existing ransomware detection systems frequently prioritize accuracy, explainability, or robustness independently rather than integrating all three into a unified and deployable framework.
\end{enumerate}

Motivated by these gaps, this study presents DDQN-MLP, a DRL-guided adaptive learning framework for behavioral ransomware detection using Windows~11 sandbox telemetry. The framework integrates five tightly coupled components: a modern Windows~11 behavioral dataset derived from ANY.RUN reports, a training-time DDQN module that selects discrete sample-importance weights, a lightweight MLP classifier deployed exclusively during inference, a post-hoc explainability module incorporating SHAP-gradient alignment analysis, and a systematic feature-space adversarial robustness evaluation with adversarial training as a defense mechanism.

This architecture introduces a decoupled training-time design that cleanly separates structural optimization from runtime deployment constraints. By restricting reinforcement learning entirely to the offline training loop, the framework avoids the computational latency and resource overhead typically associated with runtime DRL systems. Consequently, the final deployed detector remains lightweight and high-throughput while still benefiting from sophisticated sequential optimization during training.

This optimization architecture explains why value-based DRL is superior to simpler weighting strategies. Heuristic methods such as focal loss and class-weighted loss rely on static rules that remain blind to evolving model dynamics. Similarly, reactive weighting approaches adjust importance scores based only on immediate batch-level snapshots without memory mechanisms. In contrast, the DDQN maintains an experience replay buffer that captures historical optimization trajectories, stabilizing learning under severe behavioral heterogeneity. Comprehensive empirical benchmarking demonstrates that the DDQN-MLP framework achieves superior and statistically stable optimization performance compared with conventional supervised and alternative DRL-based approaches.

The primary contributions of this study are summarized as follows:

\begin{enumerate}
\item \textbf{Decoupled Training-Time DRL Sample Optimization Engine:} We formulate classifier training as a finite MDP and implement a DDQN-based offline adaptive sample-weighting controller that dynamically scales training importance based on batch-level optimization trajectories. After training, the DRL controller is completely discarded, leaving only a lightweight MLP for efficient runtime inference.

\item \textbf{High-Fidelity Modern Windows~11 Sandbox Dataset:} We construct a contemporary behavioral dataset from the ANY.RUN cloud sandbox platform containing 2,000 balanced instances across 30 ransomware families and benign applications. The resulting 103-dimensional feature space captures modern Windows~11 telemetry, including CryptoAPI interactions and kernel-level behavioral semantics.

\item \textbf{Mathematical Formalization of Explanation--Sensitivity Alignment:} We augment SHAP and LIME explainability analysis with a novel explanation-sensitivity validation mechanism based on Pearson, Spearman, and Jaccard alignment metrics between post-hoc explanations and empirical model gradients.

\item \textbf{Rigorous Adversarial Stress Testing and Hardening:} We evaluate the framework against FGSM, BIM, JSMA, and PGD white-box feature-space attacks across multiple perturbation levels and demonstrate that integrated adversarial training substantially improves robustness without degrading clean-data performance.

\item \textbf{Exhaustive Empirical Validation and Ablation Proofing:} Through a 5-fold stratified cross-validation protocol spanning 17 configurations involving statistical baselines, fixed-weighting strategies, PPO/A2C DRL variants, and both MLP and TabNet backbones, we demonstrate that DDQN-MLP achieves superior detection accuracy (99.30\%) and stable optimization under severe behavioral heterogeneity.
\end{enumerate}

The remainder of this paper is organized as follows. Section~II reviews behavioral ransomware detection, reinforcement learning for cybersecurity, explainable artificial intelligence, and adversarial robustness. Section~III presents the DDQN-MLP framework, including dataset construction, feature preprocessing, adaptive sample weighting, and joint optimization. Section~IV describes the experimental setup and evaluation protocol. Section~V presents the experimental results, including detection performance, explanation consistency analysis, adversarial robustness evaluation, and ablation studies. Section~VI discusses the findings and limitations, and Section~VII concludes the paper and outlines future research directions.

\section{Related Work}

This section reviews four core research domains: (1) ransomware behavioral analysis, (2) deep reinforcement learning for cybersecurity, (3) explainable artificial intelligence (XAI) for security analytics, and (4) adversarial machine learning for malware detection. For each domain, representative approaches are summarized, and the structural gaps motivating the proposed training-time adaptive framework are identified.

\subsection{Ransomware Behavioral Analysis}

Behavioral analysis is widely adopted in ransomware detection because static signatures fail to detect code obfuscation, polymorphism, and evolving attack workflows. Existing frameworks analyze runtime indicators such as system calls, filesystem modifications, registry updates, process interactions, and network behavior to distinguish ransomware families from benign applications~\cite{ref24,ref37,ref38}. A broad range of machine learning (ML) and deep learning (DL) models, including Convolutional Neural Networks (CNNs), Long Short-Term Memory (LSTM) networks, Hidden Markov Models (HMMs), and supervised contrastive learning techniques, have been applied to sandbox-generated behavioral telemetry with promising performance~\cite{ref8,ref39}.

Despite these advances, two major limitations remain. First, conventional behavioral detectors rely entirely on static supervised optimization strategies in which all samples contribute equally during gradient updates. This uniform optimization assumption performs poorly when behavioral data exhibit severe heterogeneity and strong class overlap. Second, most existing security benchmarks continue to rely on legacy Windows environments (Windows~7/10) and low-interaction sandboxes such as Cuckoo. These environments fail to capture modern operating-system behaviors including cloud-integrated security telemetry, modified privilege schemes, and updated CryptoAPI execution flows, creating a substantial gap between experimental evaluations and real-world enterprise threats.

\subsection{Deep Reinforcement Learning for Cybersecurity and Ransomware Detection}

The limitations of fixed supervised optimization and outdated sandbox environments highlight the need for training mechanisms capable of dynamic and context-aware sample prioritization. Reinforcement learning (RL) naturally satisfies this requirement. Unlike conventional supervised learning pipelines that commit to a static optimization objective before training, an RL controller adaptively modifies its weighting policy based on feedback from the model's current optimization state. This flexibility is particularly suitable for mitigating the behavioral heterogeneity that destabilizes conventional neural network training. Consequently, DRL has been widely applied across cybersecurity domains including intrusion detection~\cite{ref40,ref41,ref25,ref42}, phishing classification~\cite{ref28}, malware detection~\cite{ref43,ref44}, IoT monitoring~\cite{ref26,ref45}, and automated vulnerability forecasting.

However, the application of value-based and policy-based DRL within ransomware analytics remains limited. Existing frameworks primarily focus on static Portable Executable (PE) metadata~\cite{ref46}, vulnerability forecasting~\cite{ref47}, or coarse Android permission vectors~\cite{ref48}, which are either vulnerable to packing-based evasion or lack sufficient behavioral granularity for robust endpoint monitoring. More importantly, no prior study has formalized a value-based DRL agent as an offline training-time adaptive sample-weighting controller. Existing DRL-based security frameworks typically deploy the RL agent directly during runtime inference, introducing substantial processing latency and memory overhead that limit practical deployment in high-throughput endpoint environments.

\subsection{Explainability in Security Analytics and Reinforcement Learning}

Automated security systems must establish strong structural trust before deployment; therefore, improved classifier accuracy alone is insufficient if analysts cannot audit the underlying decision process. This requirement has motivated the integration of explainable artificial intelligence (XAI) into security analytics. XAI encompasses techniques that make model decisions transparent and interpretable~\cite{ref49}. Post-hoc explanation methods such as SHAP and LIME are widely used to identify feature attributions driving malware classification decisions. Furthermore, several recent studies have incorporated post-hoc explanations into DRL-driven intrusion response systems~\cite{ref33,ref34} to improve interpretability for human analysts.

Despite these advances, a fundamental theoretical limitation remains unresolved. Existing security systems do not verify whether post-hoc explanations mathematically align with the internal parameter sensitivities that govern the classifier optimization process. Consequently, explanations may appear human-plausible while failing to reflect the true decision boundaries learned by the model.

\subsection{Adversarial Robustness in Malware and Ransomware Detection}

Prior malware detection research has extensively investigated adversarial attack and defense strategies to improve model robustness~\cite{ref50,ref51}. Among adversarial attack methods, gradient-based evasion techniques such as the Fast Gradient Sign Method (FGSM), Basic Iterative Method (BIM), Jacobian Saliency Map Attack (JSMA), and Projected Gradient Descent (PGD) are widely studied because they achieve high evasion success with minimal feature perturbations~\cite{ref52}. Defensive techniques including adversarial training, defensive distillation, and ensemble hardening have improved the resilience of generic malware classifiers.

However, ransomware-specific adversarial robustness remains insufficiently explored. Existing studies primarily focus on static malware representations and rarely evaluate behavioral ransomware telemetry under adversarial conditions. More importantly, the interaction between training-time adaptive DRL sample weighting and feature-space adversarial defense mechanisms has not been formally investigated under severe behavioral heterogeneity.

\subsection{Summary and Positioning}

Existing literature has addressed behavioral ransomware detection, DRL-driven optimization, post-hoc explainability, and adversarial robustness largely as isolated research problems. No prior framework integrates these components into a unified pipeline trained on modern Windows~11 behavioral telemetry. Each dimension independently addresses one of the structural limitations identified in previous work, including static optimization objectives, low behavioral fidelity, explanation--sensitivity misalignment, and unquantified adversarial vulnerability. The proposed DDQN-MLP framework consolidates these elements into a single cohesive architecture while maintaining practical deployment feasibility through a strict separation between training-time DRL optimization and lightweight runtime inference.

\section{Methodology}
\label{sec:methodology}

The architecture consists of two interconnected tracks: (i) the proposed DDQN-MLP framework, which integrates adaptive training-time sample weighting with a lightweight inference-time classifier, and (ii) a suite of comparative baselines encompassing classical machine learning, fixed-weighting neural methods, and alternative reinforcement learning (RL) variants to rigorously contextualize the empirical advantages of the proposed framework.

\subsection{Proposed DDQN-MLP Framework}
\label{subsec:proposed_framework}

\begin{figure*}[!t]
    \centering
    \includegraphics[width=0.55\textwidth]{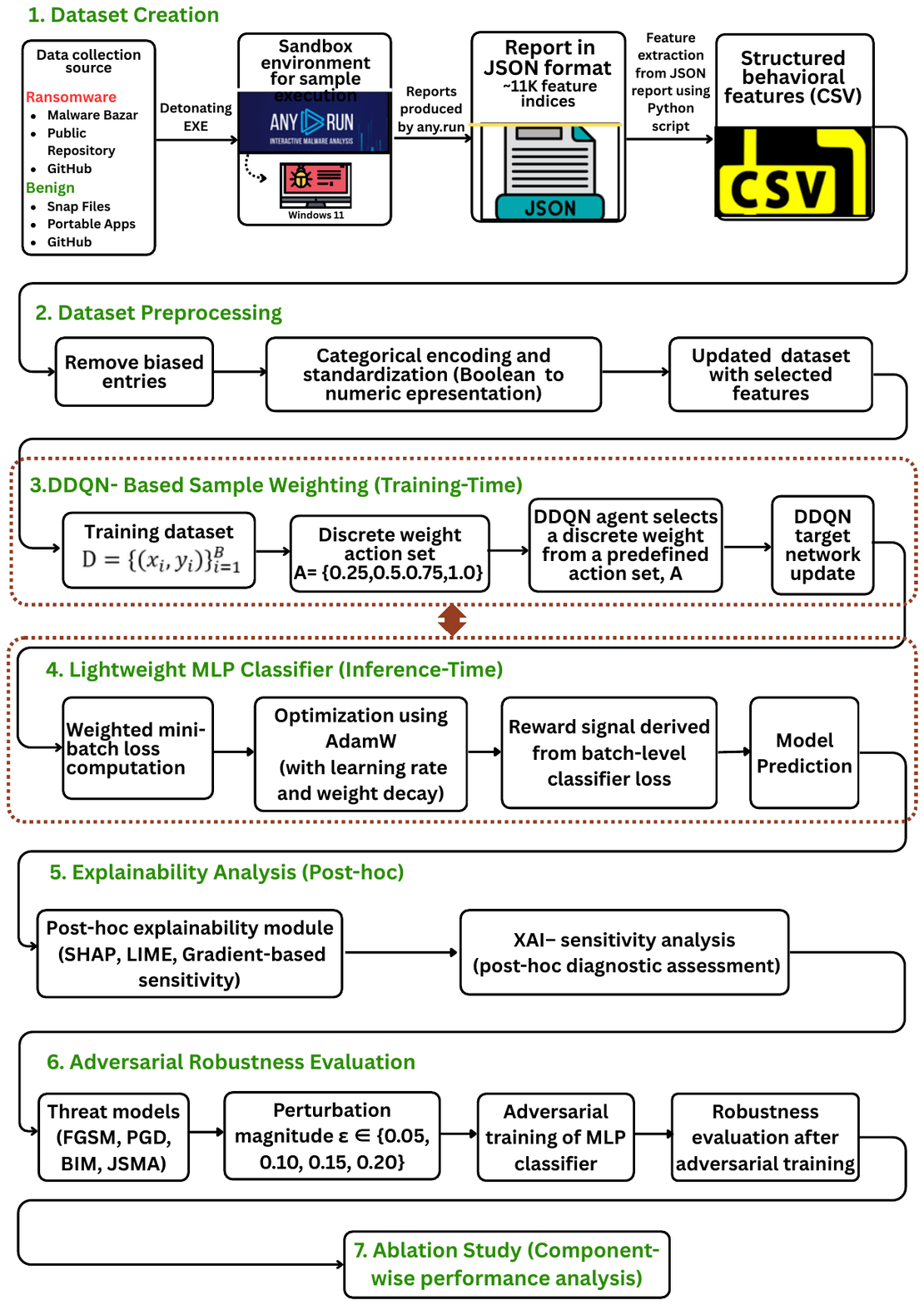}
    \caption{Overall workflow of the proposed DDQN-MLP framework for ransomware detection.}
    \label{fig:overall_workflow}
\end{figure*}

The proposed framework combines a DDQN-based adaptive sample-weighting controller with a lightweight MLP classifier for behavior-based ransomware detection using structured Windows~11 telemetry. The overall workflow consists of seven consecutive phases: (1) dataset creation, (2) data preprocessing, (3) DDQN-based adaptive sample weighting during training, (4) lightweight MLP classifier optimization using the learned sample weights, (5) post-hoc explainability analysis, (6) feature-space adversarial robustness evaluation, and (7) ablation and comparative analysis. Crucially, the reinforcement learning module is used only during training and discarded afterward, ensuring that the final deployed model is entirely stripped of all RL overhead. Fig.~\ref{fig:overall_workflow} illustrates the overall process. The following subsections describe each component.

\subsubsection{Dataset Creation}
\label{subsubsec:dataset_creation}

This section describes the construction of a Windows~11 ransomware--benign behavioral dataset and the feature-engineering pipeline used to convert sandbox telemetry into machine-learning-ready data, emphasizing modern behavioral fidelity beyond legacy Cuckoo-based datasets.

\paragraph{Data Collection and Sample Acquisition.}
To ensure contemporary behavioral fidelity and resistance to evasion tactics, a balanced dataset of 2,000 binary executable samples was constructed, comprising 1,000 legitimate applications and 1,000 ransomware instances across 30 globally prevalent ransomware families. Ransomware samples were collected from MalwareBazaar~\cite{ref53} and VirusShare~\cite{ref54}, whereas benign samples were obtained from SnapFiles~\cite{ref55}, PortableApps.com~\cite{ref56}, and GitHub~\cite{ref57} to ensure low label noise and high dataset integrity. Selection criteria were applied to both ransomware families and individual samples to ensure relevance, diversity, representativeness, and operational significance.

Two criteria were used to select ransomware families: (i) high prevalence and significant organizational impact documented in multi-source threat-intelligence reports from 2019--2024~\cite{ref58,ref59,ref60,ref61,ref62,ref63}; and (ii) verification across at least two independent reports from reputable cybersecurity vendors. This process identified prominent ransomware families such as LockBit, MedusaLocker, BlackCat, Phobos, and Conti, ensuring coverage of advanced threats.

Ransomware samples within each family were selected following the procedure in~\cite{ref11}, using three VirusTotal vendor-engine criteria. A sample was retained only if: (i) at least 45 antivirus engines classified it as malicious; (ii) at least 15 antivirus engines explicitly labelled it as ransomware; and (iii) the majority of engines, or at least ten, identified it as belonging to the same ransomware family. These VirusTotal-derived criteria were used exclusively for dataset curation and family verification during sample collection. No antivirus labels, detection counts, reputation scores, or vendor-confidence metrics were included as predictive model features or used during inference.

Table~\ref{tab:ransomware_families} shows the distribution of ransomware families and sample counts.

\begin{table}[t]
\centering
\caption{Ransomware families included in this study with their sample counts. Families were selected based on global prevalence, documented impact, and multi-vendor confirmation of classification.}
\label{tab:ransomware_families}
\scriptsize
\setlength{\tabcolsep}{3pt}
\renewcommand{\arraystretch}{1.05}
\begin{tabular}{l r l r}
\toprule
\textbf{Family} & \textbf{No.} & \textbf{Family} & \textbf{No.} \\
\midrule
LockBit & 114 & Phobos & 22 \\
GandCrab & 97 & WannaCry & 21 \\
MedusaLocker & 63 & WastedLocker & 20 \\
NetWalker & 56 & BlackMatter & 20 \\
Conti & 55 & BlackBasta & 20 \\
Babuk & 52 & Ryuk & 20 \\
Cerber & 49 & RagnarLocker & 19 \\
Maze & 47 & Mespinoza/Pysa & 19 \\
Sodinokibi/REvil & 43 & AvosLocker & 15 \\
DarkSide & 41 & Avadon & 14 \\
Dharma & 36 & MountLocker & 11 \\
Thanos & 31 & Exorcist & 11 \\
TeslaCrypt & 31 & Mallox & 10 \\
Makop & 23 & Nefilim & 10 \\
Akira & 22 & BlueSky & 8 \\
\midrule
\multicolumn{4}{c}{\textbf{Total ransomware families = 30}} \\
\multicolumn{4}{c}{\textbf{Total ransomware samples = 1000}} \\
\bottomrule
\end{tabular}
\end{table}

\paragraph{Sample Execution.}
All samples were executed in a Windows~11 environment using the ANY.RUN sandbox~\cite{ref64}, which was selected for its support of modern operating systems, rich behavioral telemetry, and improved resistance to sandbox-evasion techniques compared with legacy Cuckoo-based systems~\cite{ref65}. Each execution generated a detailed JSON report containing approximately 250 behavioral fields.

\paragraph{Feature Extraction.}
The feature-engineering pipeline converts raw ANY.RUN JSON reports into structured behavioral representations. Each JSON report was flattened into a key--value map and aggregated into semantic feature groups, including global execution metadata, filesystem activity, registry and process counters, API statistics, network behavior, dropped artifacts, and DLL activity. Incident-level behaviors and process-specific behavioral scores were also incorporated to capture fine-grained ransomware actions. This extraction process initially yielded a 105-dimensional feature vector for each sample.

\subsubsection{Data Preprocessing}
\label{subsubsec:data_preprocessing}

Prior to model training, a lightweight preprocessing stage was applied to ensure reproducibility, prevent data leakage, and enable unbiased learning. Two potentially class-biased metadata attributes, namely threat\_level and verdict\_score, were explicitly removed during preprocessing to prevent label leakage and shortcut learning. This resulted in a final 103-dimensional behavioral feature space. Additionally, Boolean-valued attributes were standardized to a numeric representation by mapping TRUE to 1 and FALSE to 0 across the feature matrix, ensuring type consistency and enabling subsequent feature scaling within the machine learning and reinforcement learning pipelines.

The detailed feature extraction logic, including JSON normalization,
behavioral aggregation, summarization into a 103-dimensional
representation, and preprocessing, is provided in Algorithm~\ref{alg:feature_extraction}.

\begin{algorithm}[t]
\caption{ANY.RUN JSON Behavioral Feature Extraction and Preprocessing}
\label{alg:feature_extraction}
\scriptsize
\begin{algorithmic}[1]
\REQUIRE Root directory $D$ containing subfolders with ANY.RUN JSON reports
\ENSURE Cleaned CSV file with one 103-dimensional numeric feature vector per sample

\STATE \textbf{function} EXTRACT\_FEATURES\_FROM\_JSON(json\_report)
\STATE $M \leftarrow$ FLATTEN\_JSON(json\_report) \COMMENT{key--value map}
\STATE $f \leftarrow$ zero vector of length 105 \COMMENT{initial raw extraction footprint}

\STATE \COMMENT{Global metadata}
\STATE $f[\mathrm{global}] \leftarrow$ EXTRACT\_GLOBAL($M$)

\STATE \COMMENT{Filesystem statistics}
\STATE $f[\mathrm{fs}] \leftarrow$ AGGREGATE\_FILESYSTEM($M$)

\STATE \COMMENT{Registry, process, and network counters}
\STATE $f[\mathrm{counters}] \leftarrow$ AGGREGATE\_COUNTERS($M$)

\STATE \COMMENT{API, dropped artifacts, and DLL activity}
\STATE $f[\mathrm{artifacts}] \leftarrow$ AGGREGATE\_ARTIFACTS($M$)

\STATE \COMMENT{Incident-level behavior}
\FOR{$i = 0$ to $19$}
    \STATE $f[\mathrm{incident\_count}_i] \leftarrow M[\mathrm{incidents}[i].\mathrm{count}]$
    \STATE $f[\mathrm{incident\_threat}_i] \leftarrow M[\mathrm{incidents}[i].\mathrm{threatLevel}]$
\ENDFOR

\STATE \COMMENT{DNS reputation scores}
\FOR{$i = 0$ to $9$}
    \STATE $f[\mathrm{dns\_rep}_i] \leftarrow M[\mathrm{dnsRequests}[i].\mathrm{reputationNumber}]$
\ENDFOR

\STATE \COMMENT{Process behavior scores}
\STATE behaviours $\leftarrow$ \{autoStart, knownThreat, stealing, executableDropped, network, debugOutput, privEscalation\}
\FOR{$p = 0$ to $4$}
    \FOR{each $b$ in behaviours}
        \STATE $f[\mathrm{procScore}(p,b)] \leftarrow M[\mathrm{processes}[100+p].\mathrm{scores.specs}[b]]$
    \ENDFOR
\ENDFOR

\STATE \textbf{return} $f$
\STATE \textbf{end function}

\STATE $F \leftarrow$ empty list
\FOR{each subfolder $s$ in $D$}
    \FOR{each JSON file $j$ in $s/json$}
        \STATE $J \leftarrow$ PARSE\_JSON($j$)
        \STATE $f \leftarrow$ EXTRACT\_FEATURES\_FROM\_JSON($J$)
        \STATE append $f$ to $F$
    \ENDFOR
\ENDFOR

\STATE WRITE\_CSV(``raw\_features.csv'', $F$)
\STATE $df \leftarrow$ READ\_CSV(``raw\_features.csv'')
\STATE Drop threat\_level and verdict\_score to prevent label leakage
\STATE $df \leftarrow df \setminus \{\mathrm{threat\_level}, \mathrm{verdict\_score}\}$

\FOR{each cell $x$ in $df$}
    \IF{$x =$ ``TRUE''}
        \STATE replace $x \leftarrow 1.0$
    \ELSIF{$x =$ ``FALSE''}
        \STATE replace $x \leftarrow 0.0$
    \ENDIF
\ENDFOR

\STATE WRITE\_CSV(``cleaned\_features.csv'', $df$)
\end{algorithmic}
\end{algorithm}

\begingroup
\setlength{\abovedisplayskip}{3pt}
\setlength{\belowdisplayskip}{3pt}
\setlength{\abovedisplayshortskip}{1.5pt}
\setlength{\belowdisplayshortskip}{1.5pt}

\subsubsection{DDQN-Based Adaptive Sample Weighting}
\label{subsubsec:ddqn_weighting}

Instead of formulating ransomware detection as an online runtime control problem, the sequential execution aspect arises from the evolving optimization trajectory of the classifier across consecutive training batches. By adaptively weighting uncertain and difficult behavioral samples, the proposed training-time strategy addresses data heterogeneity and stabilizes the optimization dynamics. DDQN is well-suited to this setting because the action space is discrete and low-dimensional, corresponding to predefined sample-weighting levels. This enables the framework to emphasize hard-to-classify samples while reducing the influence of less informative instances.

The adaptive weighting mechanism is formulated as a Markov Decision Process (MDP) and implemented using DDQN as follows.

\paragraph{State Space ($S_t$).}
At each training step $t$, the agent observes a structured state vector $S_t$ derived from batch-level training statistics. To ensure complete reproducibility, $S_t$ is explicitly defined as a concatenated vector,
\begin{equation}
S_t = [L_t, C_t]
\end{equation}

where $L_t \in \mathbb{R}^{+}$ denotes the mean batch-level classifier loss and $C_t \in [0,1]$ represents the current network prediction confidence, measured as the average distance of output probabilities from the classification threshold.

\paragraph{Action Space ($A$).}
The agent selects an action from a low-dimensional discrete action set:
\begin{equation}
A = \{0.25, 0.5, 0.75, 1.0\}
\label{eq:action_space}
\end{equation}

where each action corresponds to a weight $w_i$ applied directly to the classification loss of sample $i$, allowing the framework to dynamically scale down the influence of trivial samples while amplifying focus on high-uncertainty targets.

\paragraph{Reward Function ($r_t$).}
The reward signal driving policy convergence is defined as the negative batch-level classification loss:
\begin{equation}
r_t = -\mathcal{L}_{\mathrm{batch}}
\label{eq:reward}
\end{equation}

This reward is optimization-driven rather than behavior-driven. It does not directly encode feature relevance or class semantics; instead, it provides a scalar learning signal encouraging the agent to select weighting actions that reduce training error and improve downstream convergence.

To stabilize learning and decorrelate updates, transition tuples $(s_t, a_t, r_t, s_{t+1})$ are stored in an experience replay buffer.

To mitigate action-value overestimation, DDQN is employed instead of the standard DQN by maintaining separate online and target Q-networks. At each update step, the online network selects the greedy action:
\begin{equation}
a^\ast = \arg\max_{a} Q_{\mathrm{online}}(s_{t+1}, a)
\label{eq:greedy_action}
\end{equation}

while the target network evaluates that action to compute the temporal-difference (TD) target:
\begin{equation}
Y_t = r_t + \gamma Q_{\mathrm{target}}(s_{t+1}, a^\ast)
\label{eq:td_target}
\end{equation}

where $\gamma$ denotes the discount factor.

The TD error is optimized using the Huber (SmoothL1) loss:
\begin{equation}
\mathcal{L}_{\mathrm{TD}} =
\mathrm{SmoothL1}
\left(
Q_{\theta}(s_t,a_t), Y_t
\right)
\label{eq:td_loss}
\end{equation}

with gradient clipping applied to stabilize updates under noisy, loss-derived rewards:
\begin{equation}
\|\nabla\| \gets \min(\|\nabla\|,\mathrm{CLIP\_NORM})
\label{eq:gradient_clip}
\end{equation}

Owing to noisy and non-stationary reward signals derived from batch-level classifier loss, the DDQN double-network architecture assigns temporal credit more reliably than standard DQN or contextual bandits, which frequently overestimate action values. The target network was periodically synchronized with the online network to stabilize training convergence.

\subsubsection{Lightweight MLP Classifier}
\label{subsubsec:mlp_classifier}

To preserve a low computational footprint suitable for high-throughput endpoint deployment, the proposed framework enforces a strict separation of concerns: the DDQN guides optimization during training but is discarded after convergence, leaving only a lightweight MLP for inference-time deployment. This separation confines the computational overhead of reinforcement learning entirely to the offline training phase while maintaining an efficient deployment-time detector suitable for practical security environments.

The MLP backbone was intentionally selected for its low inference overhead, architectural simplicity, and deployment practicality, thereby isolating the contribution of the DDQN-guided adaptive weighting mechanism. Unlike heavier transformer-based or attention-driven architectures, the lightweight MLP enables efficient inference on structured telemetry while ensuring that performance improvements originate primarily from the adaptive weighting strategy rather than increased model complexity.

The classifier operates on standardized 103-dimensional behavioral feature vectors extracted from Windows~11 runtime telemetry. Let the classifier be defined as:
\begin{equation}
f_{\theta} : \mathbb{R}^{d} \rightarrow \mathbb{R}
\label{eq:mlp_mapping}
\end{equation}

where $f_{\theta}$ denotes the MLP classifier parameterized by $\theta$ and outputs a raw logit value.

The network begins with an input batch-normalization layer, followed by three fully connected layers with widths of 256, 128, and 64. Each hidden layer is coupled with batch normalization and a Sigmoid Linear Unit (SiLU) activation function, enabling stable and expressive non-linear feature learning at low computational cost. The final layer is a single-unit output layer producing raw logits. No activation function is applied to the output layer; instead, probability calibration and decision thresholding are performed externally during inference. Only the trained MLP is retained for deployment, whereas all reinforcement learning components are excluded from the final inference pipeline.

\subsubsection{Joint DDQN--MLP Training Procedure}
\label{subsubsec:joint_training}

This subsection describes how the DDQN-based adaptive weighting mechanism (Section~\ref{subsubsec:ddqn_weighting}) and MLP classifier (Section~\ref{subsubsec:mlp_classifier}) interact during training, as illustrated in Fig.~\ref{fig:joint_framework}. The DDQN agent and MLP classifier were jointly optimized using an interleaved mini-batch training process across all epochs. At each mini-batch iteration $\{(x_i,y_i)\}_{i=1}^{B}$, the DDQN agent selects a discrete sample weight $w_i$ according to the state--action policy described in Section~\ref{subsubsec:ddqn_weighting}. These weights are used to scale the classification loss, yielding the weighted objective:

\begin{equation}
\mathcal{L}_{\mathrm{weighted}}
=
\frac{1}{B}
\sum_{i=1}^{B}
w_i \cdot
\mathcal{L}
\left(
f_{\theta}(x_i), y_i
\right)
\label{eq:weighted_loss}
\end{equation}

The MLP parameters were updated through backpropagation using this weighted loss, whereas the DDQN agent was optimized independently through temporal-difference learning using replayed transitions. Consequently, the two components remain loosely coupled: the DDQN influences optimization only through loss modulation and does not directly modify the forward computation of the classifier.

After convergence, the DDQN agent, replay buffer, and Q-networks are discarded. Inference is then performed solely using the trained MLP classifier. This strict separation between training-time reinforcement learning and inference-time classification ensures efficient deployment, architectural simplicity, and reproducibility.

\begin{center}
\includegraphics[width=0.95\columnwidth]{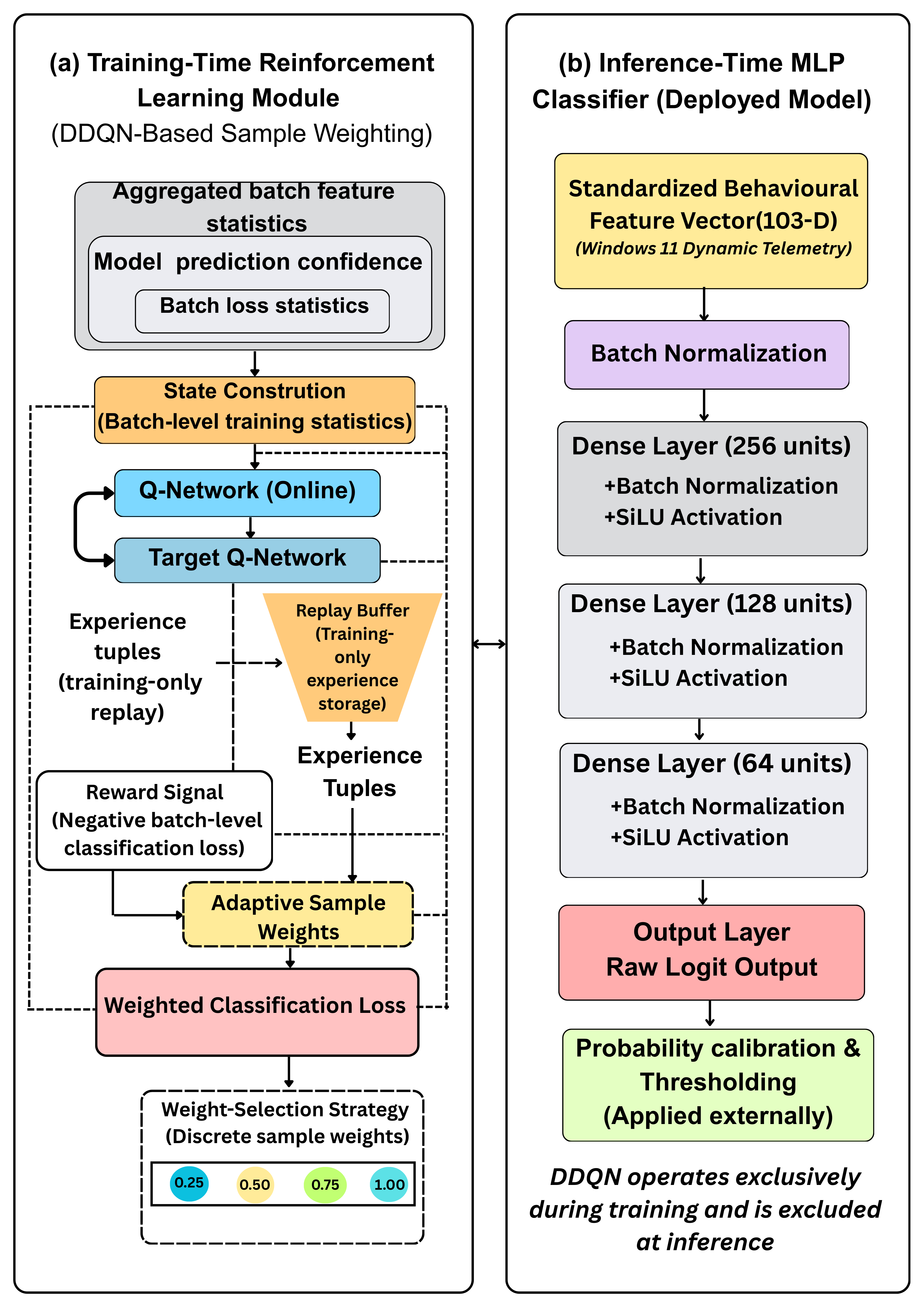}
\captionof{figure}{The proposed DDQN-guided learning framework for ransomware detection comprises two clearly separated components: (a) a training-time reinforcement learning module for adaptive sample weighting based on batch-level training statistics, and (b) a lightweight MLP classifier used exclusively during inference.}
\label{fig:joint_framework}
\end{center}
Overall, Fig.~\ref{fig:joint_framework} provides a deployment-accurate representation of the proposed DDQN-guided framework. The DDQN operates exclusively during training to influence optimization through loss weighting and is entirely excluded during deployment, whereas post-hoc explanation--sensitivity alignment analysis is applied solely for diagnostic assessment and does not affect training or inference.

The complete joint optimization workflow of the DDQN-based weighting mechanism and MLP classifier is presented in Algorithm~\ref{alg:ddqn_training}.
In Algorithm~\ref{alg:ddqn_training}, each sample-specific state is defined using batch-level training statistics and classifier confidence. Because the reward is derived at the batch level, replay storage adopts a self-transition approximation $(s_j, a_j, r_t, s_j)$ to stabilize DDQN updates.

\begin{algorithm}[t]
\caption{DDQN-Based Adaptive Sample Weighting Integrated with Mini-Batch Classifier Training}
\label{alg:ddqn_training}
\scriptsize
\begin{algorithmic}[1]

\REQUIRE Training dataset $D$, mini-batch size $B$, discrete weight action set $A$, classifier $f_\theta$, DDQN networks $(Q_{\mathrm{online}}, Q_{\mathrm{target}})$, replay buffer $M$, discount factor $\gamma$, label smoothing factor $\alpha$, gradient clipping threshold CLIP\_NORM, exploration parameters $(\epsilon_{\mathrm{init}}, \epsilon_{\mathrm{min}}, \epsilon_{\mathrm{decay}})$, target-network update interval $T_{\mathrm{sync}}$

\ENSURE Trained classifier $f_\theta$; DDQN weight-selection strategy is used only during training

\STATE \textbf{Initialization}
\STATE Initialize classifier parameters $\theta$
\STATE Initialize $Q_{\mathrm{online}}$ with random weights; set $Q_{\mathrm{target}} \leftarrow Q_{\mathrm{online}}$
\STATE Initialize replay buffer $M \leftarrow \emptyset$
\STATE Set exploration rate $\epsilon \leftarrow \epsilon_{\mathrm{init}}$

\STATE \textbf{Training Loop}

\FOR{each epoch}
    \STATE Shuffle training samples

    \FOR{each mini-batch $\mathcal{B}=\{(x_i,y_i)\}_{i=1}^{B}$}

\STATE \textbf{$\triangleright$ State construction and action selection}

        \FOR{each sample $x_i \in \mathcal{B}$}
            \STATE Construct a batch-level state $s_i$ from aggregated training statistics derived from the 103-dimensional feature space (e.g., loss and prediction confidence)
            \STATE Select action $a_i$ using $\epsilon$-greedy weight-selection strategy on $Q_{\mathrm{online}}$
            \STATE Assign sample weight $w_i \in A$ corresponding to $a_i$
        \ENDFOR

\STATE \textbf{$\triangleright$ Weighted classifier update}

        \STATE Apply label smoothing to ground-truth labels using factor $\alpha$

        \STATE Compute weighted mini-batch loss:
        $\mathcal{L}_{\mathrm{weighted}}
        =
        \frac{1}{B}
        \sum_{i=1}^{B}
        w_i \cdot
        \mathcal{L}(f_\theta(x_i), y_i)$

        \STATE Update classifier parameters $\theta$ via backpropagation on $\mathcal{L}_{\mathrm{weighted}}$

        \STATE \textbf{$\triangleright$ Reward construction}

        \STATE Set reinforcement reward $r = -\mathcal{L}_{\mathrm{batch}}$

         \STATE \textbf{$\triangleright$ Experience storage using self-transition approximation}

        \STATE Select subset $\widetilde{\mathcal{B}} \subseteq \mathcal{B}$

        \FOR{each $x_j \in \widetilde{\mathcal{B}}$}
            \STATE Store transition $(s_j, a_j, r, s_j)$ in replay buffer $M$
        \ENDFOR

        \STATE \textbf{$\triangleright$ DDQN update via experience replay}

        \IF{$|M| \geq B$}

            \STATE Sample mini-batch of transitions from $M$

            \FOR{each sampled transition tuple $(s_t,a_t,r_t,s_{t+1})$}

                \STATE Compute greedy action
                $a^\ast = \arg\max_{a} Q_{\mathrm{online}}(s_{t+1},a)$

                \STATE Compute DDQN target
                $Y_t = r_t + \gamma Q_{\mathrm{target}}(s_{t+1},a^\ast)$

            \ENDFOR

            \STATE Update $Q_{\mathrm{online}}$ by minimizing Huber loss

            \STATE Apply gradient clipping with threshold CLIP\_NORM

        \ENDIF

        \STATE \textbf{$\triangleright$ Target network synchronization}

        \IF{update step mod $T_{\mathrm{sync}} = 0$}
            \STATE Update target network parameters
            $Q_{\mathrm{target}} \leftarrow Q_{\mathrm{online}}$
        \ENDIF

        \STATE \textbf{$\triangleright$ Exploration decay}

        \STATE $\epsilon \leftarrow \max(\epsilon_{\mathrm{min}}, \epsilon \cdot \epsilon_{\mathrm{decay}})$

    \ENDFOR

\ENDFOR

\STATE \textbf{return} trained classifier $f_\theta$

\STATE Discard DDQN agent and replay buffer; deploy $f_\theta$ alone for inference

\end{algorithmic}
\end{algorithm}

\begingroup
\setlength{\abovedisplayskip}{3pt}
\setlength{\belowdisplayskip}{3pt}
\setlength{\abovedisplayshortskip}{1.5pt}
\setlength{\belowdisplayshortskip}{1.5pt}

\subsubsection{Post-hoc Explanation--Sensitivity Alignment Analysis}
\label{subsubsec:explainability}

To support transparent ransomware detection, the proposed framework integrates post-hoc explainability with SHAP--gradient alignment analysis. Post-hoc explainability was implemented using SHAP for global and local feature attributions and LIME for instance-level explanations of ambiguous and misclassified samples. Together, these techniques provide complementary population-level insights and fine-grained interpretability of the learned decision behavior.

In parallel, a gradient-based feature sensitivity proxy was computed from the trained classifier. This analysis was performed solely as a post-hoc diagnostic assessment to evaluate whether SHAP-based feature attributions remain consistent with classifier gradient-based loss sensitivity patterns, rather than as a mechanism influencing optimization or learning.

The alignment analysis was quantified using four complementary metrics: Pearson correlation for magnitude agreement, Spearman correlation for rank agreement, top-$k$ variance explained for attribution compactness, and Jaccard similarity across folds for feature-rank stability.

\paragraph{Pearson Correlation (Magnitude Agreement).}

Pearson correlation coefficient was used to quantify magnitude agreement between SHAP importance and gradient-based feature sensitivity. Let
\begin{equation}
I^{\mathrm{SHAP}}
=
[I_1^{\mathrm{SHAP}}, \ldots, I_d^{\mathrm{SHAP}}]
\label{eq:shap_vector}
\end{equation}

and
\begin{equation}
I^{\nabla}
=
[I_1^{\nabla}, \ldots, I_d^{\nabla}]
\label{eq:grad_vector}
\end{equation}

denote the vectors of mean absolute SHAP importance and gradient-based sensitivity, respectively.

Since gradient importance is undefined for non-differentiable models, alignment metrics are evaluated only over the valid feature subset:
\begin{equation}
\mathcal{M}
=
\{
k \mid I_k^{\nabla}\ \mathrm{is\ defined}
\}
\label{eq:valid_subset}
\end{equation}

Pearson correlation is then computed as
\begin{equation}
\rho_p
=
\frac{
\mathrm{cov}
(
I_{\mathcal{M}}^{\mathrm{SHAP}},
I_{\mathcal{M}}^{\nabla}
)
}{
\sigma(I_{\mathcal{M}}^{\mathrm{SHAP}})
\,
\sigma(I_{\mathcal{M}}^{\nabla})
}
\label{eq:pearson}
\end{equation}

Higher values of $\rho_p$ $(\rho_p \geq 0.80)$ indicate strong magnitude agreement between SHAP-based feature attributions and gradient-based sensitivity patterns.

\paragraph{Spearman Rank Correlation (Ordering Consistency).}

Spearman rank correlation was used to evaluate ordering consistency between SHAP importance and gradient-based sensitivity:
\begin{equation}
\rho_s
=
\rho_p
(
\mathrm{rank}(I_{\mathcal{M}}^{\mathrm{SHAP}}),
\mathrm{rank}(I_{\mathcal{M}}^{\nabla})
)
\label{eq:spearman}
\end{equation}

Higher values of $\rho_s$ $(\rho_s \geq 0.75)$ indicate stable feature prioritization under potentially non-linear relationships.

\paragraph{Variance Explained by Top-$k$ Features (Compactness).}

Explanation compactness was quantified as the proportion of total SHAP attribution mass captured by the top-$k$ ranked features:
\begin{equation}
V_k
=
\frac{
\sum_{i \in \tau_k}
I_i^{\mathrm{SHAP}}
}{
\sum_{j=1}^{d}
I_j^{\mathrm{SHAP}}
}
\label{eq:variance_explained}
\end{equation}

where $\tau_k$ denotes the index set of the top-$k$ features ranked by SHAP importance. Higher values of $V_k$ $(V_{20} \geq 0.85)$ indicate increasingly compact explanations with attribution concentrated on a limited subset of dominant features.

\paragraph{Feature Rank Stability (Jaccard Similarity Across Folds).}

To evaluate explanation robustness and reproducibility across cross-validation folds, feature-rank stability was measured using Jaccard similarity between top-$k$ feature sets:
\begin{equation}
J_{p,q}
=
\frac{
|\tau_k^{(p)} \cap \tau_k^{(q)}|
}{
|\tau_k^{(p)} \cup \tau_k^{(q)}|
}
\label{eq:jaccard_pair}
\end{equation}

The average cross-fold Jaccard similarity is then computed as
\begin{equation}
J
=
\frac{
2
}{
F(F-1)
}
\sum_{p<q}
J_{p,q}
\label{eq:jaccard_avg}
\end{equation}

where $F$ denotes the number of folds and $\tau_k^{(p)}$ represents the top-$k$ SHAP feature set in fold $p$. Greater explanation stability is indicated by an average Jaccard similarity of at least $0.70$.

Collectively, these metrics evaluate whether SHAP-based explanations remain aligned with classifier decision-sensitivity patterns linked to the loss signal guiding DDQN-based sample weighting. The adopted thresholds assess explanation agreement and stability consistent with established practices in correlation analysis, attribution compactness, and explanation robustness evaluation~\cite{ref66,ref67}. The resulting explanations demonstrate decision consistency, stability, and reproducibility without influencing model optimization or guaranteeing policy correctness.

\subsubsection{White-Box Feature-Space Adversarial Evaluation}
\label{subsubsec:adversarial_eval}

We evaluated the resilience of the inference-time MLP against adversarial manipulation in the feature space. Because the proposed framework operates on structured runtime telemetry rather than raw executable binaries, feature-space perturbations provide a controlled and operationally relevant robustness evaluation. Since the DDQN functions solely as a training-time optimization controller and is discarded after convergence, the deployed attack surface is restricted entirely to the lightweight MLP inference module.

\paragraph{Adversarial Threat Model.}

A white-box feature-space threat model was adopted in which the adversary possesses full access to classifier parameters and gradient information. Four gradient-based attacks were considered: Fast Gradient Sign Method (FGSM), Basic Iterative Method (BIM), Projected Gradient Descent (PGD), and Jacobian-based Saliency Map Attack (JSMA).

FGSM applies a single-step perturbation using the loss gradient, BIM iteratively refines FGSM-style updates, PGD performs bounded iterative optimization and is widely regarded as one of the strongest first-order adversarial attacks, whereas JSMA perturbs only highly salient features using forward-pass sensitivity analysis, making it structurally distinct from gradient-sign-based attacks. All attacks directly perturb the 103-dimensional behavioral feature vector to induce classifier misclassification.

\paragraph{Adversarial Training (AT).}

To improve robustness, adversarial training was employed as the primary defense mechanism. As illustrated in Fig.~\ref{fig:adv_pipeline}, the evaluation consists of two conditions.

In the \emph{before-AT} condition, the classifier was trained exclusively on clean training data and subsequently evaluated under adversarial attack. In the \emph{after-AT} condition, the classifier was retrained using clean training samples combined with adversarial examples generated exclusively at $\epsilon = 0.10$ from the training set. The test set remained entirely unseen during retraining to ensure a fair and leakage-free evaluation protocol.

The retrained classifier was then evaluated against all four attacks across perturbation magnitudes
\[
\epsilon \in \{0.05, 0.10, 0.15, 0.20\}
\]
thereby enabling direct quantification of the robustness improvements provided by adversarial training.

\begin{figure*}[t]
\centering
\includegraphics[width=0.95\textwidth]{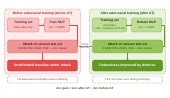}
\caption{White-box feature-space adversarial evaluation pipeline showing two conditions: before adversarial training (AT), where the MLP is trained using clean data only, and after adversarial training, where the MLP is retrained using adversarial examples generated exclusively at $\epsilon=0.10$ from the training data. The same unseen test set was used under both conditions to ensure a leakage-free comparison.}
\label{fig:adv_pipeline}
\end{figure*}

\subsection{Comparative Baselines}
\label{subsubsec:comparative_baselines}

This subsection describes the models used exclusively for comparative evaluation and ablation analysis and does not constitute part of the proposed framework. To isolate the contribution of DDQN-based adaptive sample weighting, alternative reinforcement learning agents were evaluated by replacing the DDQN module while preserving the same state representation, action space, reward formulation, and classifier backbone.

The evaluated reinforcement learning baselines included:

\begin{itemize}
\setlength{\itemsep}{1pt}
\setlength{\parskip}{0pt}
\setlength{\parsep}{0pt}

\item \textbf{Deep Q-Network (DQN):} A vanilla value-based agent that learns action values through temporal-difference updates without explicit overestimation control.

\item \textbf{Advantage Actor--Critic (A2C):} An actor--critic approach that jointly optimizes a stochastic policy and value function to reduce variance.

\item \textbf{Proximal Policy Optimization (PPO):} A policy-gradient method that stabilizes learning through clipped surrogate objectives.

\item \textbf{Random Weighting Agent:} A non-learning baseline that assigns sample weights uniformly at random from the predefined action set.
\end{itemize}

For non-RL comparison, the baseline MLP and TabNet were retained as the primary comparators. The baseline MLP uses the same architecture as the proposed classifier but is trained without RL-guided weighting, whereas TabNet serves as a representative attentive tabular-learning model.

Additionally, to assess whether the performance gains of DDQN-guided adaptive weighting are justified relative to simpler weighting strategies, several additional baselines were evaluated under identical experimental settings. These included a Class-Weighted MLP, which applies inverse class-frequency sample weights using the same classifier architecture and therefore provides a direct comparison against learned adaptive weighting through fixed static weighting; a Focal-Loss MLP trained with focal loss $(\gamma = 2)$ to down-weight easy samples using a predefined weighting schedule; Logistic Regression; Linear SVM; RBF SVM; KNN; and Gaussian Na\"ive Bayes. Collectively, these baselines establish comparative performance bounds across fixed-weighting neural methods, lightweight classical classifiers, and alternative learning paradigms, enabling rigorous assessment of whether DDQN-guided adaptive weighting provides measurable advantages beyond simpler weighting mechanisms.

\section{Experimental Setup}
\label{sec:experimental_setup}

This section outlines the experimental design used to evaluate the proposed DDQN--MLP framework on a high-fidelity Windows 11 behavioral dataset, emphasizing reproducibility and rigorous assessment through stratified cross-validation. Table~\ref{tab:model_config} summarizes the model architecture and training configuration, whereas Table~\ref{tab:evaluation_protocol} details the evaluation protocol, including the explainability and adversarial robustness analyses.

\vspace{2mm}

\noindent
\captionof{table}{Model Configuration (DDQN--MLP)}
\label{tab:model_config}

{\scriptsize

\setlength{\tabcolsep}{3pt}
\renewcommand{\arraystretch}{1.05}

\begin{tabularx}{\columnwidth}{@{}p{0.30\columnwidth}X@{}}
\toprule
\textbf{Component} & \textbf{Specification} \\
\midrule
Dataset & 2,000 Windows 11 samples: 1,000 ransomware and 1,000 benign \\
Sandbox & Windows 11 ANY.RUN \\
Features & 103 behavioral features \\
DDQN State & Batch-level loss and confidence from the 103-D feature space \\
DDQN Actions & Sample weights $\{0.25,0.5,0.75,1.0\}$ \\
DDQN Parameters & $\gamma=0.40$, $\epsilon:1.0\rightarrow0.05$, decay = 0.992, replay memory = 10k \\
DDQN Training & Batch size = 64, Adam ($lr=1\times10^{-3}$), clip = 5.0 \\
MLP Architecture & $103\rightarrow256\rightarrow128\rightarrow64\rightarrow1$ \\
Activations & SiLU + BatchNorm \\
Regularization & Dropout: 0.25 / 0.20 / 0.10 \\
Loss Function & Weighted BCE via DDQN action $\alpha(a_t)$ \\
MLP Optimizer & AdamW, $lr=2\times10^{-3}$, OneCycleLR \\
Label Smoothing & 0.02 \\
\bottomrule
\end{tabularx}

}

\vspace{3mm}

\noindent
\captionof{table}{Evaluation Protocol}
\label{tab:evaluation_protocol}

{\scriptsize

\setlength{\tabcolsep}{3pt}
\renewcommand{\arraystretch}{1.05}

\begin{tabularx}{\columnwidth}{@{}p{0.30\columnwidth}X@{}}
\toprule
\textbf{Aspect} & \textbf{Setting} \\
\midrule
Platform & Google Colab Pro CPU-only, Intel Xeon, 51 GB RAM \\
Software & Python 3.10, PyTorch 2.1, sklearn 1.3, SHAP 0.44, LIME 0.2.0, pytorch-tabnet \\
Validation & 5-fold stratified cross-validation: 80\% train / 20\% validation \\
Training & 25 epochs, early stopping with patience = 6 \\
Metrics & Accuracy, Precision, Recall, F1-score, ROC--AUC \\
Explainability & SHAP with 150 background and evaluation samples; LIME top-10 features \\
XAI Alignment & Pearson/Spearman SHAP--gradient correlation, top-20 variance, Jaccard stability \\
Adversarial Attacks & FGSM, BIM, JSMA, PGD; $\epsilon=0.05$ to $\epsilon=0.20$ \\
Robustness Metrics & Accuracy before/after AT, accuracy drop, accuracy gain \\
Runtime Analysis & Training time, inference time, RL overhead \\
\bottomrule
\end{tabularx}

}

\section{Experimental Results}

With the experimental framework established, results were analyzed across four dimensions.
First, classification performance was evaluated to determine if DDQN-guided adaptive
weighting improved baselines. Second, SHAP--gradient alignment assesses model transparency
and consistency. Third, white-box adversarial evaluation examines model robustness under
feature-space attacks. Finally, ablation analysis isolates individual design components to
verify observed gains from architectural choices.

\subsection{Overall Classification Performance}

Table~\ref{tab:cv_results} reports the cross-validated performance (mean $\pm$ std) of
all evaluated configurations, including nine non-RL baselines and eight RL-augmented
variants using DQN, DDQN, A2C, and PPO across both the MLP and TabNet backbones.

\begin{table*}[!t]
\centering
\caption{Cross-validation performance summary (mean $\pm$ standard deviation across five folds) for all evaluated baseline, classical, and RL-augmented models}
\label{tab:cv_results}

\footnotesize
\setlength{\tabcolsep}{4pt}
\renewcommand{\arraystretch}{1.05}

\begin{tabular}{lccccccc}
\toprule
\textbf{Model} & \textbf{Accuracy} & \textbf{Precision} & \textbf{Recall} & \textbf{F1-score} & \textbf{ROC--AUC} & \textbf{Train (s)} & \textbf{Pred (s)} \\
\midrule
Baseline MLP & 0.9880$\pm$0.0033 & 0.9919$\pm$0.0025 & 0.9840$\pm$0.0058 & 0.9879$\pm$0.0034 & 0.9983$\pm$0.0016 & 10.65 & 0.02 \\
MLP Class-Weighted & 0.9895$\pm$0.0037 & 0.9950$\pm$0.0050 & 0.9840$\pm$0.0065 & 0.9894$\pm$0.0037 & 0.9992$\pm$0.0008 & 2.95 & 0.002 \\
MLP Focal Loss & 0.9890$\pm$0.0034 & 0.9970$\pm$0.0028 & 0.9810$\pm$0.0065 & 0.9889$\pm$0.0034 & 0.9989$\pm$0.0019 & 4.20 & 0.002 \\
Baseline TabNet & 0.9825$\pm$0.0088 & 0.9840$\pm$0.0102 & 0.9810$\pm$0.0086 & 0.9825$\pm$0.0088 & 0.9964$\pm$0.0025 & 206.39 & 0.37 \\
Logistic Regression & 0.9845$\pm$0.0051 & 0.9939$\pm$0.0056 & 0.9750$\pm$0.0100 & 0.9843$\pm$0.0052 & 0.9917$\pm$0.0081 & 0.04 & 0.001 \\
Linear SVM & 0.9870$\pm$0.0033 & 0.9910$\pm$0.0055 & 0.9830$\pm$0.0067 & 0.9869$\pm$0.0033 & 0.9945$\pm$0.0049 & 0.16 & 0.002 \\
RBF SVM & 0.9825$\pm$0.0025 & 0.9830$\pm$0.0026 & 0.9820$\pm$0.0057 & 0.9825$\pm$0.0025 & 0.9987$\pm$0.0004 & 0.37 & 0.017 \\
KNN & 0.9335$\pm$0.0095 & 0.9989$\pm$0.0025 & 0.8680$\pm$0.0192 & 0.9287$\pm$0.0109 & 0.9865$\pm$0.0029 & 0.00 & 0.020 \\
Gaussian NB & 0.9360$\pm$0.0084 & 0.9792$\pm$0.0069 & 0.8910$\pm$0.0192 & 0.9329$\pm$0.0095 & 0.9852$\pm$0.0046 & 0.00 & 0.001 \\
DQN + MLP & 0.9920$\pm$0.0029 & 0.9960$\pm$0.0038 & 0.9880$\pm$0.0024 & 0.9920$\pm$0.0029 & 0.9993$\pm$0.0005 & 31.35 & 0.02 \\
\textbf{DDQN + MLP (Proposed)} & \textbf{0.9930$\pm$0.0024} & \textbf{0.9970$\pm$0.0025} & \textbf{0.9890$\pm$0.0049} & \textbf{0.9930$\pm$0.0025} & \textbf{0.9991$\pm$0.0009} & 32.47 & 0.02 \\
A2C + MLP & 0.9905$\pm$0.0037 & 0.9940$\pm$0.0058 & 0.9870$\pm$0.0024 & 0.9905$\pm$0.0037 & 0.9991$\pm$0.0008 & 174.62 & 0.02 \\
PPO + MLP & 0.9895$\pm$0.0019 & 0.9920$\pm$0.0024 & 0.9870$\pm$0.0024 & 0.9895$\pm$0.0019 & 0.9991$\pm$0.0006 & 114.77 & 0.02 \\
DQN + TabNet & 0.9890$\pm$0.0020 & 0.9930$\pm$0.0060 & 0.9850$\pm$0.0032 & 0.9890$\pm$0.0020 & 0.9993$\pm$0.0004 & 27.26 & 0.01 \\
DDQN + TabNet & 0.9880$\pm$0.0024 & 0.9900$\pm$0.0031 & 0.9860$\pm$0.0037 & 0.9880$\pm$0.0025 & 0.9990$\pm$0.0012 & 26.20 & 0.01 \\
A2C + TabNet & 0.9880$\pm$0.0024 & 0.9930$\pm$0.0051 & 0.9830$\pm$0.0024 & 0.9879$\pm$0.0024 & 0.9987$\pm$0.0010 & 113.20 & 0.01 \\
PPO + TabNet & 0.9885$\pm$0.0021 & 0.9929$\pm$0.0040 & 0.9840$\pm$0.0025 & 0.9889$\pm$0.0020 & 0.9988$\pm$0.0012 & 109.35 & 0.01 \\
\bottomrule
\end{tabular}
\end{table*}

The results demonstrate that the proposed DDQN--MLP framework achieved the strongest
overall mean performance, with the highest F1-score ($0.9930 \pm 0.0025$) and accuracy
($0.9930 \pm 0.0024$), while maintaining a low fold-to-fold variance. Across both
backbones, value-based RL methods (DQN and DDQN) generally outperform policy gradient
methods (A2C and PPO), indicating more stable optimization of difficult behavioral samples.

\textbf{Comparison with Baseline and Weighted-Loss Models:} Relative to the Baseline MLP,
DDQN+MLP improves the F1-score by $+0.51\%$, precision by $+0.82\%$, and recall modestly,
while also showing stronger fold-to-fold stability. It further achieves a $+0.36\%$ F1
gain over the class-weighted MLP and $+0.41\%$ gain over the focal-loss MLP. Although
these improvements are modest, they consistently favor the DDQN-guided approach across the
cross-validation folds. Unlike class-weighted and focal-loss schemes, which apply fixed
weighting rules, the DDQN agent learns a state-conditioned policy that adapts to the
current classifier state.

\textbf{Comparison with TabNet and Classical Baselines:} Baseline TabNet achieves
competitive accuracy but incurs substantially higher training cost (206.39~s) without
improving the F1-score over the stronger MLP-based variants. Among the classical baselines,
Linear SVM achieved the highest F1-score ($0.9869 \pm 0.0033$), followed by Logistic
Regression ($0.9843 \pm 0.0052$) and RBF SVM ($0.9825 \pm 0.0025$). KNN and Gaussian
Na\"{i}ve Bayes performed substantially worse, particularly in terms of recall. These
results indicate that the performance advantage of DDQN+MLP remains evident relative to
both the neural and classical baselines.

In addition, DDQN+MLP achieved near-perfect discrimination (ROC-AUC $= 0.9991 \pm 0.0009$)
with minimal inference latency, highlighting its practical suitability as an efficient
ransomware detection model.

\textbf{Confusion Matrix:} As shown in Fig.~\ref{fig:confusion_matrix}, the confusion matrix
indicates a reliable classification performance for the DDQN+MLP model. The model
correctly classified $49.85\%$ and $49.45\%$ of the benign and ransomware samples,
respectively, yielding an overall accuracy of $99.30\%$. Misclassification rates
remained minimal, with only $0.15\%$ of benign samples incorrectly labeled as ransomware
and $0.55\%$ of ransomware samples misclassified as benign. These results indicate a strong
discriminative capability and balanced detection performance across the cross-validation
folds.

\begin{center}
\includegraphics[width=0.60\columnwidth]{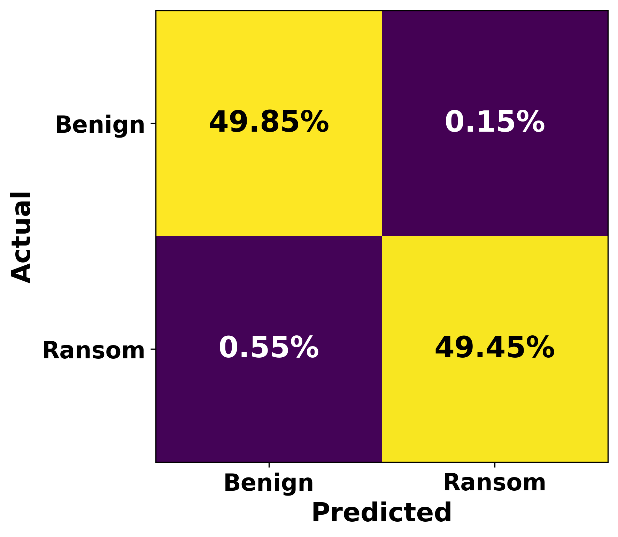}
\captionof{figure}{Confusion matrix of the proposed DDQN+MLP framework.}
\label{fig:confusion_matrix}
\end{center}

\textbf{ROC Curve:} Fig.~\ref{fig:roc_curve} shows the ROC curve of the proposed DDQN-guided
MLP detector under the 5-fold stratified cross-validation protocol. The curve remained
close to the upper-left corner, indicating high true-positive rates at very low
false-positive rates across thresholds. The corresponding ROC-AUC of $0.9991$ reflects a
near-perfect ranking between benign and ransomware samples, confirming that the learned
scoring function provides highly reliable discrimination under in-distribution evaluation.
The dashed diagonal line represents the random-guessing baseline.

\begin{center}
\includegraphics[width=0.70\columnwidth]{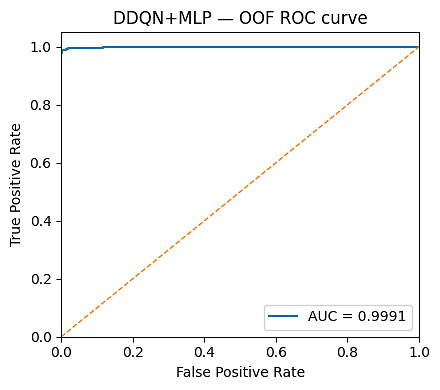}
\captionof{figure}{ROC curve of the proposed DDQN-guided MLP framework.}
\label{fig:roc_curve}
\end{center}

\textbf{Action Selection Behavior of the DDQN Agent:} Fig.~\ref{fig:action_distribution} illustrates
the distribution of discrete sample-weight actions selected by the DDQN agent during
training. The histogram shows a clear preference for an intermediate weight of $0.5$,
indicating that the agent most often assigns moderate importance to samples rather than
uniformly emphasizing or suppressing them. A weight of $1.0$ was selected less frequently
but remained substantial, suggesting that the agent selectively amplifies informative or
difficult samples when beneficial. In contrast, lower-weight actions occur relatively
infrequently. Overall, this distribution indicates that the learned policy is adaptive and
nontrivial, balancing sample emphasis and training stability rather than collapsing to a
fixed or random weighting rule.

\begin{center}
\includegraphics[width=0.88\columnwidth]{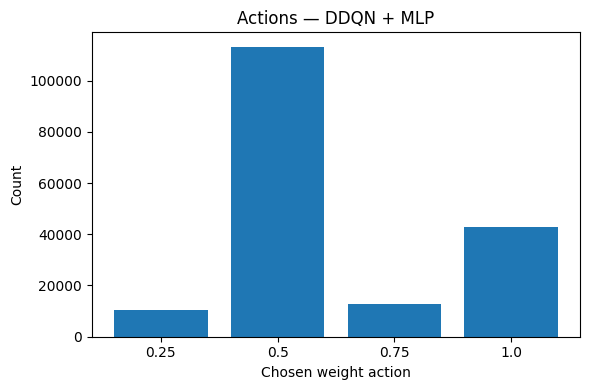}
\captionof{figure}{Action selection behavior of the DDQN agent during training.}
\label{fig:action_distribution}
\end{center}

% ============================================================
\subsection{Explainability and SHAP--Gradient Alignment Analysis}
% ============================================================

This section presents the explainability results for the DDQN--MLP model, showing that the
detection decisions are driven by a relatively small set of behaviorally meaningful features,
as evidenced by the global SHAP rankings and local SHAP/LIME explanations for
representative and misclassified samples. The SHAP--Gradient alignment analysis provides a
post-hoc diagnostic assessment of explanation consistency with classifier sensitivity
patterns and is presented as a trustworthiness-oriented diagnostic.

\subsubsection{SHAP-Based Explainability (Global + Local)}
Table~\ref{tab:shap_top20} and the SHAP bar plot in Fig.~\ref{fig:shap_top20} highlight the
top-20 features driving the DDQN-driven MLP classifier.

\begin{table*}[!h]
\centering
\caption{Top-20 Global SHAP Feature Importance for DDQN--MLP}
\label{tab:shap_top20}
\resizebox{\textwidth}{!}{%
\begin{tabular}{c l c c l c}
\hline
\textbf{Rank} & \textbf{Feature} & \textbf{Mean $|$SHAP$|$} & \textbf{Rank} & \textbf{Feature} & \textbf{Mean $|$SHAP$|$} \\
\hline
1  & proc\_anomalous\_behavior\_count & 0.447212 & 11 & proc3\_anomalous\_pattern\_flag & 0.175289 \\
2  & runtime\_6\_event\_severity & 0.395913 & 12 & runtime\_16\_event\_severity & 0.169913 \\
3  & fs\_modified\_file\_max\_size\_bytes & 0.378685 & 13 & runtime\_8\_event\_severity & 0.142028 \\
4  & runtime\_0\_event\_severity & 0.285627 & 14 & runtime\_2\_event\_severity & 0.126591 \\
5  & proc2\_anomalous\_pattern\_flag & 0.260711 & 15 & runtime\_17\_event\_severity & 0.081026 \\
6  & runtime\_3\_event\_severity & 0.249875 & 16 & api\_file\_drop\_count & 0.077986 \\
7  & runtime\_9\_event\_severity & 0.244844 & 17 & runtime\_13\_event\_severity & 0.076839 \\
8  & runtime\_5\_event\_severity & 0.208234 & 18 & dns\_endpoint\_risk\_profile\_1 & 0.066012 \\
9  & runtime\_1\_event\_severity & 0.191472 & 19 & runtime\_12\_event\_severity & 0.062263 \\
10 & runtime\_4\_event\_severity & 0.179581 & 20 & runtime\_18\_event\_severity & 0.061769 \\
\hline
\end{tabular}
}
\end{table*}

\begin{center}
\includegraphics[width=0.98\columnwidth]{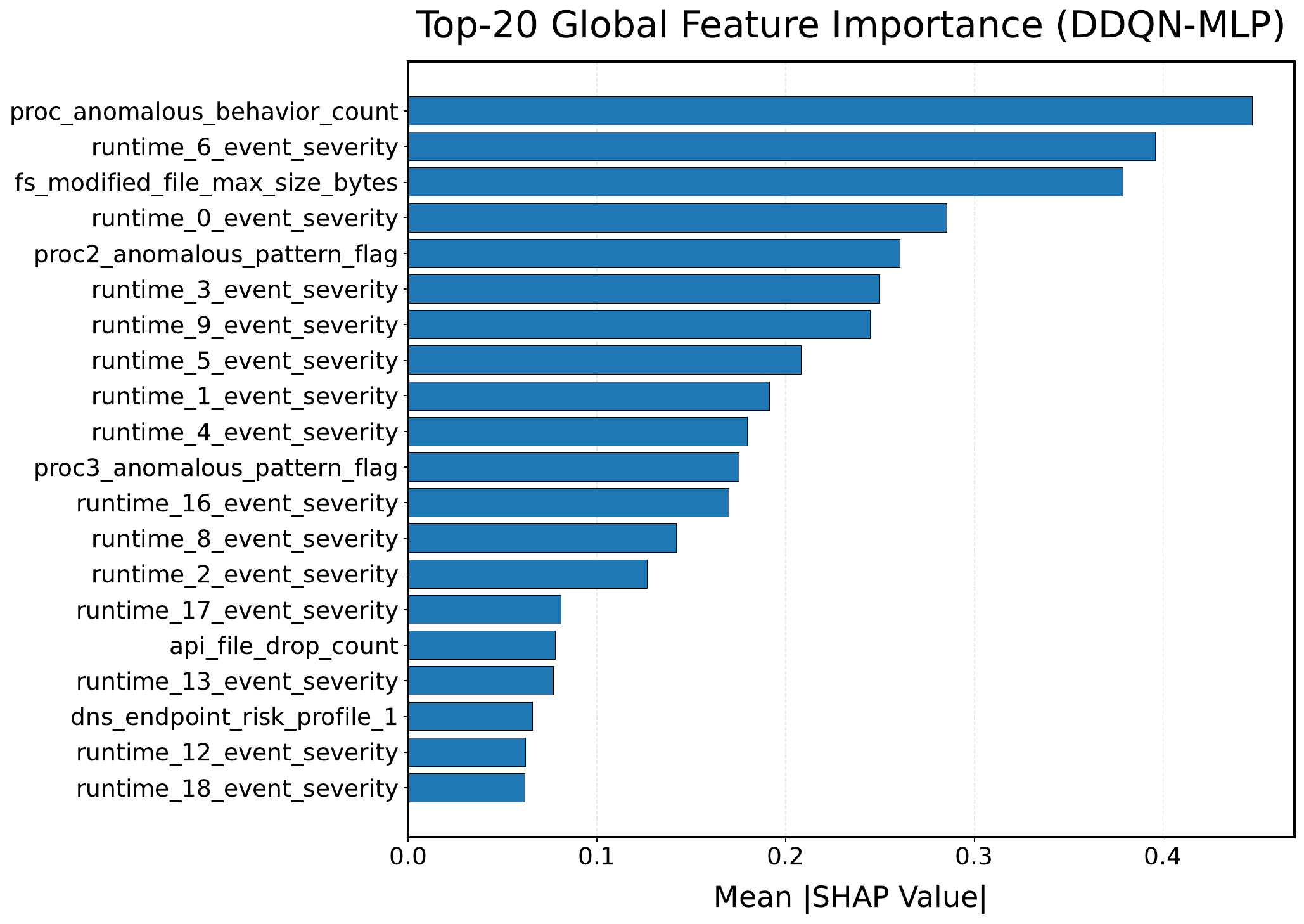}
\captionof{figure}{Top-20 global feature importance for the DDQN--MLP classifier based on mean absolute SHAP values.}
\label{fig:shap_top20}
\end{center}

The most influential features correspond to the core ransomware behaviors, including anomalous process activity (\texttt{proc \_anomalous \_behavior \_count}), file-system modifications (\texttt{fs \_modified \_file \_max \_size \_bytes}), and multiple runtime event-severity indicators (e.g., \texttt{runtime \_0 \_event \_severity} and \texttt{proc2 \_anomalous \_pattern \_flag}). The dominance of these behavioral features indicates that the model relies on meaningful ransomware cues rather than dataset artifacts or spurious correlations. The rapid decline in SHAP magnitude beyond the top-ranked features suggests a compact decision structure in which dominant behavioral signals account for most predictive power. This compactness supports interpretability, stability, and forensic reliability, which are essential for security-critical deployment.

\paragraph{SHAP Beeswarm Analysis for Global Feature Attributions.}
The SHAP beeswarm plot in Fig.~\ref{fig:shap_beeswarm} illustrates the global feature
influence across the validation samples. Key behavioral indicators ---
\texttt{proc\_anomalous\_behavior\_count}, \texttt{runtime\_6\_event\_severity},
\texttt{fs\_modified\_file\_max\_size\_bytes}, and \texttt{runtime\_0\_event\_severity} ---
exhibited high attribution magnitudes with broad dispersion, indicating consistent
contributions rather than isolated effects. Notably, \texttt{proc2\_anomalous\_pattern\_flag}
and \texttt{proc3\_anomalous\_pattern\_flag} displayed a wide positive SHAP spread at high
feature values, reflecting their strong discriminative role when anomalous process patterns
were elevated. The clear separation between the positive and negative SHAP values further
demonstrates that the model leverages distinct behavioral signatures to differentiate
ransomware from benign activities. These results confirm that DDQN-guided sample weighting
produces stable global decision patterns based on meaningful behavioral data.

\begin{center}
    \includegraphics[width=0.95\columnwidth]{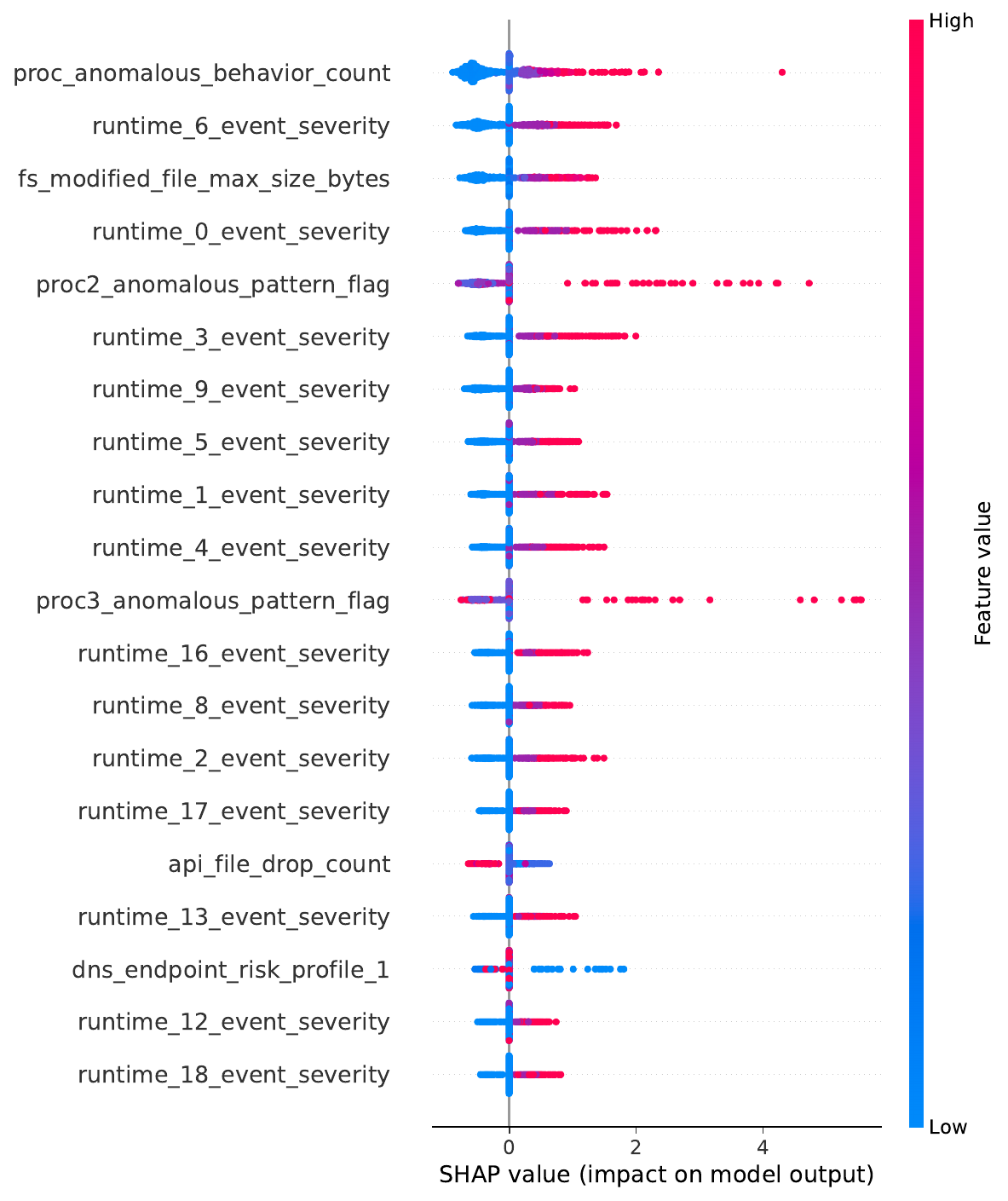}
    \captionof{figure}{Global SHAP beeswarm plot of the DDQN--MLP classifier. Each point represents
    a single validation sample, with the color indicating the feature value magnitude
    (red = high, blue = low). The plot illustrates the direction and dispersion of feature
    contributions across the validation set, confirming that
    \texttt{proc\_anomalous\_behavior\_count}, \texttt{runtime\_6\_event\_severity},
    \texttt{fs\_modified\_file\_max\_size\_bytes}, and
    \texttt{proc2\_anomalous\_pattern\_flag} consistently drive predictions toward the
    ransomware class at high feature values.}
    \label{fig:shap_beeswarm}
\end{center}

\paragraph{SHAP Waterfall Analysis (Local Explanation).}
The SHAP waterfall plot in Fig.~\ref{fig:shap_waterfall} explains an individual prediction by decomposing the model output into additive feature contributions. For the representative benign instance $(f(x)=-5.201)$, runtime event severity indicators, including \texttt{runtime\_6\_event\_severity}, \texttt{runtime\_0\_event\_severity}, \texttt{runtime\_4\_event\_severity}, \texttt{runtime\_5\_event\_severity}, and \texttt{runtime\_1\_event\_severity}, together with file-system activity (\texttt{fs\_modified\_file\_max\_size\_bytes}) and anomalous process behavior (\texttt{proc\_anomalous\_behavior\_count}, \texttt{proc2\_anomalous\_pattern\_flag}), collectively drive the prediction strongly toward the benign class. In contrast, network-related features such as \texttt{net\_http\_request\_count}, \texttt{net\_dns\_query\_count}, and registry operation indicators (\texttt{reg\_ops\_read}, \texttt{reg\_ops\_write}) contribute only marginally in the positive direction. These observations indicate that the final decision is dominated by a compact set of behavioral indicators characterized by suppressed runtime severity and limited file-system activity. The instance-level explanation remains consistent with the global SHAP analysis and confirms that individual predictions are behaviorally grounded rather than driven by incidental correlations.

\begin{center}
\includegraphics[width=0.85\columnwidth]{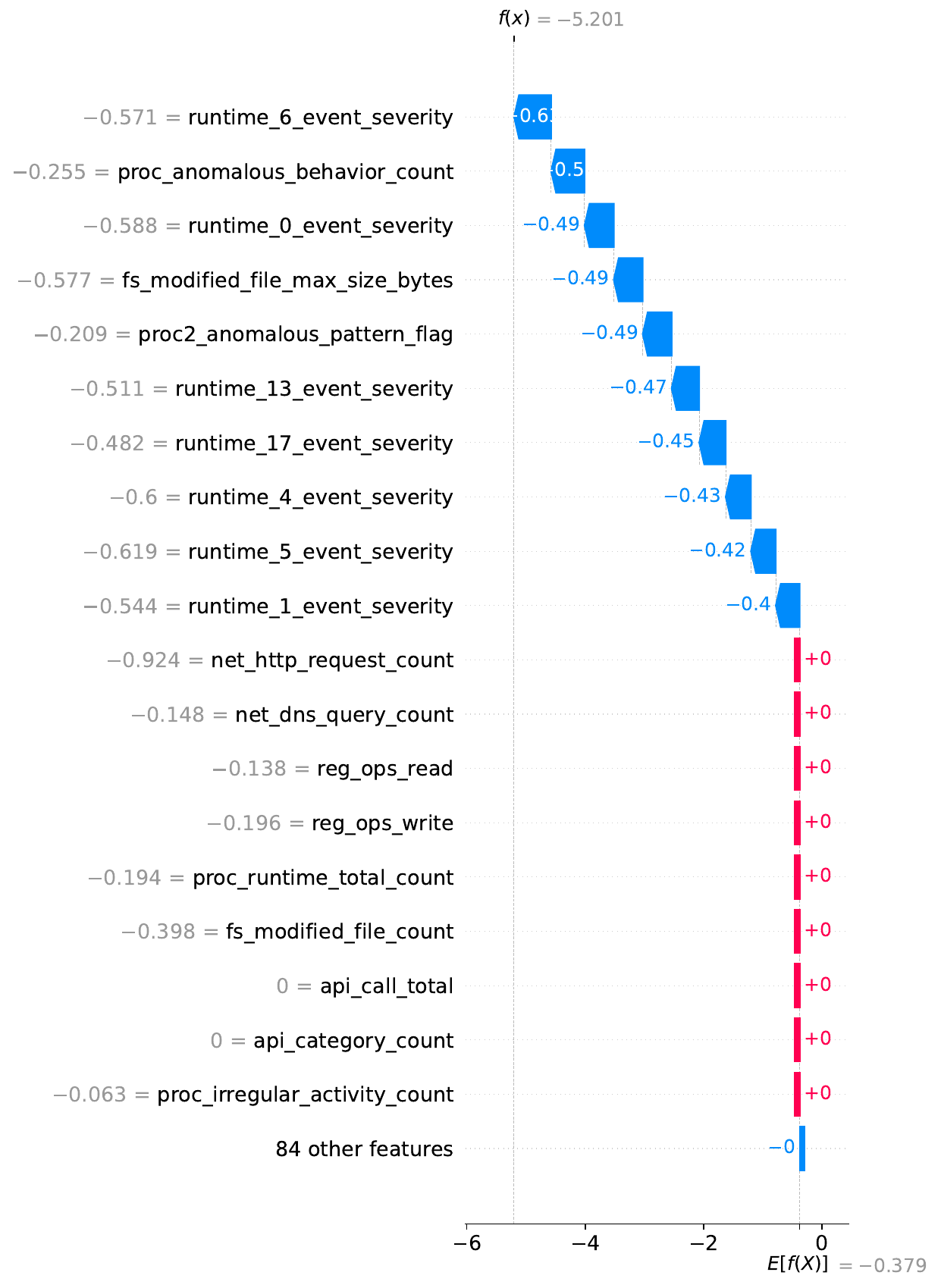}
\captionof{figure}{Local SHAP waterfall plot for a representative benign instance correctly classified by the DDQN--MLP classifier. Each bar represents the additive contribution of an individual feature to the final prediction output $(f(x)=-5.201)$, starting from the base value $E[f(X)] = -0.379$. Runtime event severity indicators and file-system activity features collectively drive the prediction strongly toward the benign class.}
\label{fig:shap_waterfall}
\end{center}

\paragraph{LIME-Based Local Explanation of Misclassified Samples.}
Figs.~\ref{fig:lime_fn} and~\ref{fig:lime_fp} illustrate the LIME-based local explanations for two representative misclassified instances, one ransomware and one benign, to investigate the local decision behavior of the classifier under challenging classification conditions.

For the misclassified ransomware sample in Fig.~\ref{fig:lime_fn}, although the sample was ransomware, the DDQN--MLP classifier predicted it as benign with a probability of 0.87, while assigning only 0.13 probability to the ransomware class. The blue contributions represent features supporting the benign prediction, whereas the orange contributions support the ransomware class. This explanation shows that benign-oriented indicators such as \texttt{proc3\_anomalous\_pattern\_flag}, \texttt{fs\_modified\_file\_max\_size\_bytes}, and \texttt{runtime\_2\_event\_severity} strongly influenced the model toward the benign decision boundary. In particular, \texttt{runtime\_2\_event\_severity} exhibited a large positive value (2.60), which substantially contributed to the benign prediction. Conversely, ransomware-associated behaviors, including \texttt{runtime\_0\_event\_severity}, \texttt{runtime\_4\_event\_severity}, \texttt{runtime\_7\_event\_severity}, \texttt{runtime\_1\_event\_severity}, and \texttt{proc0\_file\_drop\_behavior}, pushed the prediction toward the ransomware class but were insufficient to outweigh the dominant benign-oriented contributions.

\begin{figure*}[!t]
\centering
\includegraphics[width=0.88\textwidth]{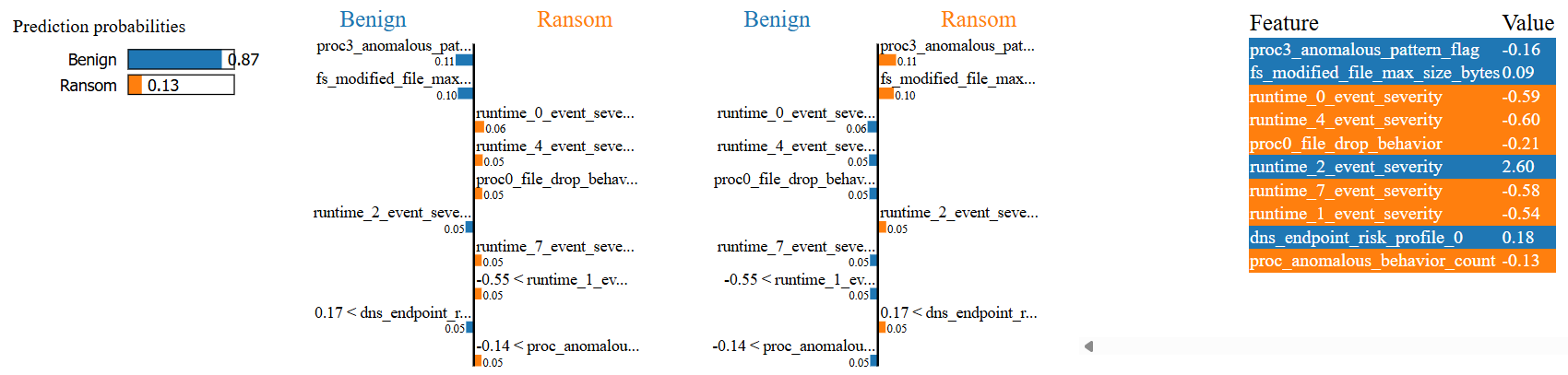}
\caption{LIME-based local explanation of a ransomware sample misclassified as benign. The figure shows the top 10 features and predicted probabilities for the benign (87\%) and ransomware (13\%) classes. The blue and orange bars indicate benign and ransomware predictions, respectively. The right panel lists the feature values for the misclassified sample, where \texttt{runtime\_2\_event\_severity} (2.60) was the dominant benign contributor, outweighing ransomware indicators and resulting in a false-negative prediction.}
\label{fig:lime_fn}
\end{figure*}

For the misclassified benign sample in Fig.~\ref{fig:lime_fp}, the opposite behavior was observed. Although the sample was genuinely benign, the classifier predicted it as ransomware with a probability of 0.95, while assigning only 0.05 probability to the benign class. This explanation reveals that ransomware-oriented indicators, including \texttt{runtime\_2\_event\_severity}, \texttt{proc3\_anomalous\_pattern\_flag}, \texttt{runtime\_0\_event\_severity}, and \texttt{proc3\_autostart\_behavior}, strongly pushed the prediction toward the ransomware decision boundary. Among these, \texttt{runtime\_2\_event\_severity} showed the strongest contribution, indicating that the benign application exhibited runtime characteristics highly similar to suspicious ransomware-like behavior. In contrast, benign-supporting features such as \texttt{api\_file\_drop\_count}, \texttt{proc\_anomalous\_behavior\_count}, and several lower-severity runtime indicators attempted to counterbalance the prediction but were insufficient to overcome the dominant ransomware-oriented contributions.

\begin{figure*}[!t]
\centering
\includegraphics[width=0.70\textwidth]{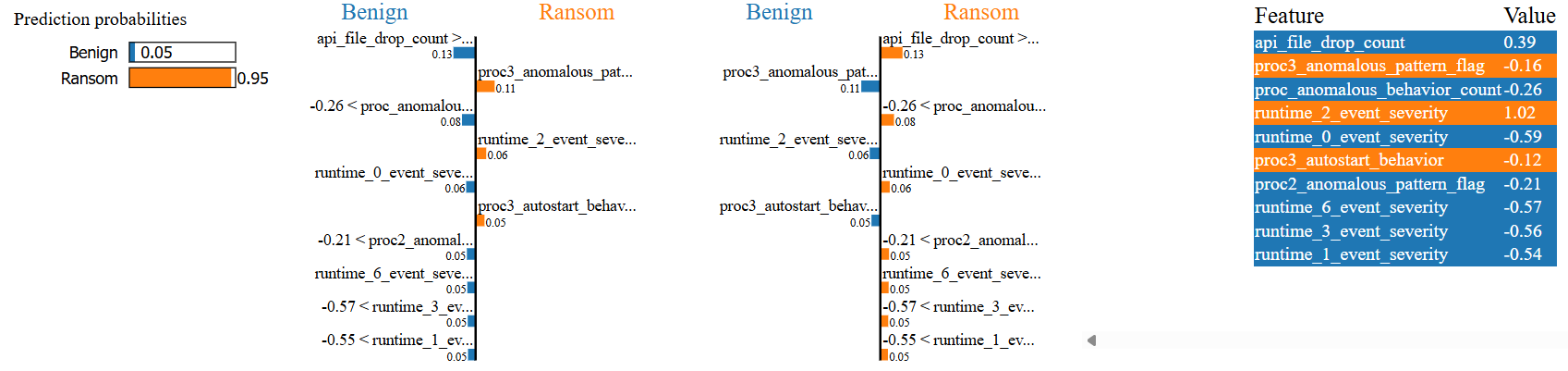}
\caption{LIME-based local explanation of a benign sample misclassified as ransomware. The figure shows the top 10 features and predicted probabilities for benign (5\%) and ransomware (95\%) classes. The orange and blue bars indicate features supporting ransomware and benign predictions, respectively. The right panel lists the feature values for the misclassified sample, showing how ransomware-like runtime behaviors in the benign application led to a high-confidence false-positive prediction.}
\label{fig:lime_fp}
\end{figure*}

Taken together, Figs.~\ref{fig:lime_fn} and~\ref{fig:lime_fp} demonstrate that the observed misclassifications were not random prediction failures but resulted from substantial behavioral overlap between sophisticated ransomware samples and complex benign applications. The LIME explanations further confirm that the DDQN--MLP classifier relied primarily on meaningful runtime behavioral indicators when making decisions, whereas the remaining classification errors emerged from ambiguous dynamic behaviors that blurred the distinction between malicious and legitimate software activity in real-world environments.

\subsubsection{SHAP--Gradient Alignment Analysis}
\label{subsec:shap_gradient_alignment}

Table~\ref{tab:shap_alignment} summarizes the quantitative SHAP--gradient alignment
metrics. A Pearson correlation of $0.8923$ ($>0.80$ target) indicated a strong global
magnitude agreement between the SHAP importance and gradient-based sensitivity. The
Spearman rank correlation of $0.8304$ ($>0.75$ target) indicates a stable feature
prioritization ordering. The top-20 SHAP features accounted for $83.32\%$ of the total
attribution mass, indicating a substantial concentration of explanatory importance within a
small feature subset. Feature rank stability across folds (Jaccard $= 0.7036$, $>0.70$
target) confirmed reproducible explanation patterns.

\vspace{6pt}
\noindent\begin{minipage}{\columnwidth}
\renewcommand{\arraystretch}{1.3}
\setlength{\tabcolsep}{4pt}
\footnotesize
\captionof{table}{SHAP--Gradient Alignment Metrics for the DDQN--MLP Model (post-hoc
diagnostic assessment).}
\label{tab:shap_alignment}
\begin{tabularx}{\columnwidth}{@{} p{2.4cm} p{1.0cm} p{1.1cm} X @{}}
\toprule
\textbf{Metric} & \textbf{Target} & \textbf{Obtained} & \textbf{Interpretation} \\
\midrule
Pearson Correlation (SHAP vs.\ Gradient)
  & $>0.80$ & $0.8923$ & Strong magnitude agreement \\
Spearman Rank Correlation
  & $>0.75$ & $0.8304$ & Consistent feature ordering \\
Variance Explained (Top-20 SHAP)
  & $>0.85$ & $0.8332$ & High attribution concentration \\
Feature Rank Stability (Jaccard)
  & $>0.70$ & $0.7036$ & Stable explanation patterns across folds \\
\bottomrule
\end{tabularx}
\end{minipage}

\vspace{6pt}
Overall, these results provide a post-hoc consistency assessment, showing that the
generated explanations are compact, stable, and broadly aligned with the classifier
sensitivity patterns associated with loss-guided sample weighting. Importantly, this
analysis is diagnostic only and does not imply the verification of policy correctness or
influence model training.

\subsection{White-Box Feature-Space Adversarial Robustness Results}
\label{subsec:adv_results}

The explainability results in Section~V.B confirm that the classifier decisions are grounded in behaviorally meaningful features; however, a model identifying important features can still remain vulnerable to adversaries that deliberately perturb those features. This evaluation therefore examines how reliably the classifier makes decisions when inputs are crafted to deceive it. Two training conditions of the same MLP classifier were compared, as illustrated in Table~\ref{tab:adv_results}. The \emph{before-AT} model was trained exclusively on clean data and evaluated directly on adversarially generated samples to establish a vulnerability baseline. The \emph{after-AT} model was retrained using clean training data combined with adversarial examples generated exclusively at $\epsilon=0.10$ from the training set (clean accuracy = 0.9925), while the test set remained entirely unseen during retraining. This ensured that the improved robustness reflected the effectiveness of the defense mechanism rather than exposure to test data. The retrained model was subsequently evaluated against all four attacks across all $\epsilon$ values ($0.05$, $0.10$, $0.15$, $0.20$), with the accuracy-gain column quantifying the direct improvement:
\[
\text{Acc gain} = \text{Acc after AT} - \text{Acc before AT}
\]

Table~\ref{tab:adv_results} reports the results across four attacks (FGSM, BIM, JSMA, and PGD) and four perturbation magnitudes. Without defense, the accuracy dropped to near-random levels across all attacks and perturbation magnitudes. After adversarial training, the accuracy was restored above 0.95 at $\epsilon=0.20$ and above 0.98 at $\epsilon=0.10$, while preserving clean-data performance. Under JSMA, adversarial training additionally yielded a marginal improvement over the clean baseline, suggesting a beneficial regularization effect against sparse saliency-based perturbations. Overall, the results confirm that the lightweight MLP combined with adversarial training achieves strong feature-space robustness with negligible clean-accuracy degradation, which is practically important for deployable ransomware detection systems.

\begin{table*}[!t]
\centering
\caption{White-box feature-space adversarial robustness of the deployed MLP classifier before and after adversarial training across four attacks (FGSM, BIM, JSMA, and PGD) and four perturbation magnitudes. The drop values are relative to the clean baseline for each condition.}
\label{tab:adv_results}

\scriptsize
\setlength{\tabcolsep}{5pt}
\renewcommand{\arraystretch}{1.08}

\begin{tabular}{lcccccc}
\toprule
\textbf{Attack} & \boldmath$\epsilon$ & \textbf{Acc before AT} & \textbf{Drop before AT} & \textbf{Acc after AT} & \textbf{Drop after AT} & \textbf{Acc gain} \\
\midrule

Clean (no attack) & --   & 0.9945 & --      & 0.9925 & --      & -- \\

FGSM & 0.05 & 0.4930 & $-0.5015$ & 0.9910 & $-0.0015$ & $+0.4980$ \\
FGSM & 0.10 & 0.5000 & $-0.4945$ & 0.9865 & $-0.0060$ & $+0.4865$ \\
FGSM & 0.15 & 0.5000 & $-0.4945$ & 0.9755 & $-0.0170$ & $+0.4755$ \\
FGSM & 0.20 & 0.5000 & $-0.4945$ & 0.9590 & $-0.0335$ & $+0.4590$ \\

BIM & 0.05 & 0.4590 & $-0.5355$ & 0.9910 & $-0.0015$ & $+0.5320$ \\
BIM & 0.10 & 0.4270 & $-0.5675$ & 0.9865 & $-0.0060$ & $+0.5595$ \\
BIM & 0.15 & 0.3955 & $-0.5990$ & 0.9735 & $-0.0190$ & $+0.5780$ \\
BIM & 0.20 & 0.3705 & $-0.6240$ & 0.9525 & $-0.0400$ & $+0.5820$ \\

JSMA & 0.05 & 0.9690 & $-0.0255$ & 0.9940 & $+0.0015$ & $+0.0250$ \\
JSMA & 0.10 & 0.5455 & $-0.4490$ & 0.9955 & $+0.0030$ & $+0.4500$ \\
JSMA & 0.15 & 0.5000 & $-0.4945$ & 0.9965 & $+0.0040$ & $+0.4965$ \\
JSMA & 0.20 & 0.5000 & $-0.4945$ & 0.9975 & $+0.0050$ & $+0.4975$ \\

PGD & 0.05 & 0.4670 & $-0.5275$ & 0.9910 & $-0.0015$ & $+0.5240$ \\
PGD & 0.10 & 0.4590 & $-0.5355$ & 0.9865 & $-0.0060$ & $+0.5275$ \\
PGD & 0.15 & 0.4595 & $-0.5350$ & 0.9730 & $-0.0195$ & $+0.5135$ \\
PGD & 0.20 & 0.4685 & $-0.5260$ & 0.9520 & $-0.0405$ & $+0.4835$ \\

\bottomrule
\end{tabular}
\end{table*}

\subsection{Contextual Ablation Study}
\label{subsec:ablation_study}

To assess the contribution of individual components in the proposed DDQN--MLP framework, we conducted a contextual ablation study across reinforcement learning, weighting strategy, classifier backbone, and adversarial defense, as summarized in Table~\ref{tab:ablation}.

\begin{table*}[!t]
\centering
\caption{Contextual Ablation of the Proposed DDQN--MLP Framework}
\label{tab:ablation}

\scriptsize
\setlength{\tabcolsep}{5pt}
\renewcommand{\arraystretch}{1.08}

\begin{tabularx}{\textwidth}{@{}p{0.18\textwidth}p{0.22\textwidth}p{0.10\textwidth}p{0.12\textwidth}X@{}}
\toprule
\textbf{Component Changed} & \textbf{Configuration} & \textbf{Metric} & \textbf{Result} & \textbf{Interpretation} \\
\midrule

Full model 
& DDQN--MLP (Proposed) 
& Acc/F1 
& 0.9930 / 0.9930 
& Best CV stability with moderate training overhead. \\

No RL weighting 
& Baseline MLP 
& Acc/F1 
& 0.9880 / 0.9879 
& Strong baseline, consistently below RL-guided weighting. \\

Fixed class weighting 
& MLP Class-Weighted 
& Acc/F1 
& 0.9895 / 0.9894 
& Fixed weighting improves the plain MLP but remains below adaptive DDQN weighting. \\

Fixed loss shaping 
& MLP Focal Loss 
& Acc/F1 
& 0.9890 / 0.9889 
& Intermediate improvement using a fixed weighting schedule rather than a learned policy. \\

Replace DDQN $\rightarrow$ DQN 
& DQN--MLP 
& Acc/F1 
& 0.9920 / 0.9920 
& DDQN overestimation control provides a small but consistent advantage. \\

Change backbone 
& DDQN--TabNet 
& Acc/F1 
& 0.9880 / 0.9880 
& Heavier architecture with lower accuracy and higher computational cost. \\

Remove learning signal 
& Random-Weight MLP 
& Acc/F1 
& 0.9895 / 0.9895 
& Learned prioritization, rather than stochastic weighting, drives the observed gains. \\

Attack model 
& DDQN--MLP + FGSM ($\epsilon=0.10$) 
& Accuracy 
& 0.5000 
& Near-random accuracy under feature-space perturbation without defense. \\

Enable defense 
& DDQN--MLP + Adversarial Training 
& Accuracy 
& 0.9865 
& Feature-space robustness restored above 98.6\% while preserving clean accuracy (0.9925). \\

\bottomrule
\end{tabularx}

\end{table*}

Removing reinforcement learning by reverting to the Baseline MLP reduced the F1-score from 0.9930 to 0.9879, indicating that adaptive weighting improves cross-validation performance and training stability. The comparison with the class-weighted MLP (F1 = 0.9894) and focal-loss MLP (F1 = 0.9889) is particularly informative: both fixed-weighting baselines improved over the plain MLP, yet neither matched the DDQN+MLP performance. This suggests that state-conditioned adaptive prioritization provides a consistent, albeit modest, advantage over fixed weighting schedules. Replacing the DDQN policy with random weighting yielded F1 = 0.9895, which remained close to the class-weighted MLP, indicating that the gain originated from learned sample prioritization rather than stochastic perturbation of the loss. Replacing DDQN with DQN slightly reduced the performance (F1 = 0.9920), suggesting that DDQN overestimation control provides a small but consistent benefit.

Changing the backbone from MLP to TabNet further reduced the performance in the RL-guided setting (DDQN--TabNet: F1 = 0.9880) while increasing the computational cost, indicating that greater architectural complexity does not improve behavioral feature learning in this setting. For reference, the non-RL Baseline TabNet performed even lower (F1 = 0.9825), reinforcing the suitability of the lightweight MLP backbone for the proposed framework. The robustness-related ablation further shows that the undefended DDQN--MLP model is highly vulnerable to FGSM attacks, with accuracy dropping to near-random levels (0.5000). However, adversarial training restored the accuracy to 0.9865 without harming clean-data performance. These results confirm that adversarial training substantially improves feature-space robustness within the DDQN--MLP framework.

\subsection{Comparison with Existing Relevant Methods}

Table~\ref{tab:comparison} compares the proposed DDQN--MLP framework with representative
ransomware detection methods reported in the literature. Because these studies differ in
operating systems, platforms, feature modalities, and evaluation protocols, the comparison
is intended as contextual positioning rather than direct experimental replication.

Existing dynamic analysis methods based on recurrent neural networks
(RNN)~\cite{ref8}, convolutional neural networks (CNN)~\cite{ref9}, regularized logistic
regression (RLR)~\cite{ref39}, and Markov modeling (MM) combined with random forests
(RF)~\cite{ref10}, achieve competitive accuracy, but they rely on fixed supervised training
and generally do not consider adaptive sample weighting, post-hoc explanation consistency,
or adversarial robustness evaluation. Other behavioral classifiers, such as RLR-based
systems~\cite{ref38}, exhibit similar limitations.

RL-based ransomware detectors in static settings, including A2C-driven
models~\cite{ref48} and DDQN applied to portable executable headers~\cite{ref46}, show the
potential of RL; however, they are limited to static feature representations, involve
higher training overhead, and do not evaluate post-hoc explanation behavior or adversarial
robustness.

In contrast, the proposed framework combines DDQN-guided adaptive sample weighting with
Windows 11 behavioral telemetry data collected from ANY.RUN. In the contextual comparison
in Table~\ref{tab:comparison}, it achieves the highest reported accuracy ($99.30\%$) with
low training and prediction latency. Unlike prior methods, the proposed framework
incorporates SHAP and LIME-based interpretability, post-hoc SHAP--gradient alignment
analysis for diagnostic assessment, and systematic feature-space adversarial robustness
evaluation with adversarial training. The resulting low false-positive ($0.15\%$) and
false-negative ($0.55\%$) rates further indicate a strong operational reliability.

Overall, Table~\ref{tab:comparison} suggests that the proposed framework uniquely combines
adaptive RL-guided training, modern behavioral telemetry, interpretability assessment,
adversarial robustness analysis, and efficient deployment within a single ransomware
detection pipeline, whereas existing studies typically address only a subset of these
aspects.

\begin{table*}[ht]
\caption{Contextual comparison of the proposed DDQN--MLP framework with representative
ransomware detection methods from the literature. This comparison is not a direct
experimental benchmark because prior studies differ in terms of datasets, operating
systems, feature modalities, and evaluation protocols.}
\label{tab:comparison}
\renewcommand{\arraystretch}{1.3}
\setlength{\tabcolsep}{3pt}
\footnotesize
\begin{tabularx}{\textwidth}{@{}
    >{\raggedright\arraybackslash}p{1.4cm}
    >{\raggedright\arraybackslash}X
    >{\centering\arraybackslash}p{0.55cm}
    >{\raggedright\arraybackslash}X
    >{\raggedright\arraybackslash}X
    >{\raggedright\arraybackslash}X
    >{\centering\arraybackslash}p{1.1cm}
    >{\centering\arraybackslash}p{1.0cm}
    >{\centering\arraybackslash}p{0.9cm}
    >{\centering\arraybackslash}p{0.9cm}
    >{\centering\arraybackslash}p{0.9cm}
@{}}
\toprule
\textbf{Ref.} &
\textbf{Analysis Technique} &
\textbf{RL} &
\textbf{Explainability} &
\textbf{Robustness evaluation} &
\textbf{Classifier} &
\textbf{Accuracy (\%)} &
\textbf{Train (s)} &
\textbf{Pred (s)} &
\textbf{FP (\%)} &
\textbf{FN (\%)} \\
\midrule
\cite{ref8}  & Dynamic & No  & Yes  & No  & RNN       & 94.26 & 182.6  & 0.43  & 0.85   & ---  \\
\cite{ref9}  & Dynamic & No  & ---  & No  & CNN       & 98.2  & ---    & ---   & 0.047  &      \\
\cite{ref39} & Dynamic & No  & No   & No  & RLR       & 96.3  & ---    & ---   & 0.0161 & ---  \\
\cite{ref10} & Dynamic & No  & No   & No  & MM + RF   & 97.28 & ---    & ---   & 4.8    & 1.5  \\
\cite{ref38} & Dynamic & No  & No   & No  & RLR       & 97.33 & ---    & ---   & ---    & ---  \\
\cite{ref48} & Static  & Yes & ---  & No  & DRL (A2C) & 89.78 & 168.27 & 0.76  & ---    & ---  \\
\cite{ref46} & Static  & Yes & ---  & No  & DDQN      & 97.9  & 120    & 0.017 & ---    & ---  \\
\midrule
\textbf{Our} \newline
\textbf{method} \newline
\textbf{(DDQN} \newline
\textbf{+MLP)} &
\textbf{Dynamic (Windows 11 behavioural telemetry)} &
\textbf{Yes} &
\textbf{SHAP + LIME} &
\textbf{FGSM/BIM/ JSMA/PGD + adversarial training} &
\textbf{DDQN-guided MLP} &
\textbf{99.30} &
\textbf{32.47} &
\textbf{0.02} &
\textbf{0.15} &
\textbf{0.55} \\
\bottomrule
\end{tabularx}
\end{table*}

\section{Discussion}

The proposed DDQN--MLP framework achieved an accuracy of $99.30\%$, an F1-score of
$0.9930$, and an ROC-AUC of $0.9991$ under 5-fold stratified cross-validation, offering
an effective and lightweight approach for ransomware detection in modern Windows 11
telemetry. Improvements across all folds suggest that DDQN-guided adaptive sample
weighting enhances the management of behavioral heterogeneity by prioritizing uncertain
samples during training. Unlike fixed strategies, the DDQN adapts the weighting based on
optimization states, resulting in stable learning. These findings are crucial because
modern ransomware often mimics legitimate high-I/O and system-management activities,
causing behavioral overlap with benign software~\cite{ref4,ref24,ref37}.

Across both backbones, value-based RL methods (DQN and DDQN) consistently outperformed
policy-gradient methods (A2C and PPO), indicating that temporal credit assignment through
experience replay is suitable for noisy batch-level reward signals~\cite{ref46}. Comparing
with fixed-weighting alternatives is also informative. The class-weighted and focal-loss
MLPs improved over the baseline but did not match the DDQN+MLP performance. The key
difference is that fixed-weighting schemes use predefined rules, whereas DDQN learns a
state-conditioned policy from the current loss and confidence statistics. This adaptivity
is beneficial when the ransomware behavior overlaps with benign high-I/O
activity~\cite{ref23,ref40,ref41,ref42}. However, the performance gap is modest, and the
class-weighted MLP is a strong practical alternative when simplicity is preferred. The
reward signal $r = -\mathcal{L}_{\text{batch}}$ is optimization-driven, guiding the
classifier toward greater discriminability in regions of behavioral confusion, without
handcrafted domain rules.

The explainability analysis showed that the classifier relied on meaningful runtime
indicators, such as runtime-event severity, anomalous process behavior, filesystem
modification, and dropped-file operations, which are consistent with known ransomware
behaviors reported in prior studies~\cite{ref5,ref8,ref38,ref39}. The post-hoc SHAP--gradient
alignment analysis checks explanation reliability by assessing whether SHAP attributions
align with classifier sensitivity patterns, addressing a limitation of earlier studies
that did not evaluate explanation consistency with model behavior~\cite{ref33,ref34}. Strong
alignment scores suggest that the explanations are grounded in the model's decision
behavior. Misclassification analysis revealed that remaining false positives and negatives
arose from behavioral overlap between sophisticated ransomware and aggressive benign
applications rather than unstable decision behavior.

The adversarial results provide an additional practical insight: despite near-perfect
clean-data accuracy, the undefended classifier fell to near-random performance under FGSM,
BIM, and PGD attacks at all tested magnitudes, showing that clean-trained models remain
vulnerable to white-box attacks, consistent with prior adversarial machine learning
research~\cite{ref50,ref51,ref52}. However, adversarial training restored accuracy above $0.95$
at $\varepsilon = 0.20$ and above $0.98$ at $\varepsilon = 0.10$ across all attacks,
whereas clean-data accuracy was fully preserved at $0.9925$. Under the JSMA, adversarial
training also yields a slight improvement over the clean baseline, suggesting a beneficial
regularization effect. These results show that substantial feature-space robustness can be
achieved without changing the inference-time model architecture, which is a practically
important finding for lightweight deployment. However, the evaluation is limited to
feature-space white-box perturbations, and problem-space adversarial testing remains
important for future studies.

The ablation results further show that the performance gain arises mainly from adaptive
sample prioritization rather than architectural complexity. Replacing DDQN with random
weighting or fixed-loss schemes reduced performance, whereas substituting the MLP backbone
with TabNet increased the computational cost without an accuracy benefit. These comparisons
confirm that each design choice in the framework is justified and contributes to
performance.

These findings have broader implications for the design of ransomware detectors. The
training-time RL separation principle improves accuracy without inference overhead,
addressing practical deployment concerns for endpoint security. The Windows 11 ANY.RUN
dataset fills a documented gap in behavioral corpora~\cite{ref8,ref9,ref30,ref31,ref32} and
provides a contemporary evaluation benchmark for 30 ransomware families. The
SHAP--gradient alignment analysis (Pearson $0.8923$, Spearman $0.8304$, Jaccard $0.7036$)
offers a novel diagnostic dimension absent from prior RL-based explainability
studies~\cite{ref33,ref34}, confirming that explanations are grounded in model sensitivity
patterns from DDQN-guided training rather than post-hoc artifacts~\cite{ref49,ref66}.

This study has several limitations. First, the dataset was moderate in size, and the
evaluation was limited to in-distribution cross-validation without family hold-out or
zero-day testing. Second, the adversarial analysis is limited to feature-space
perturbations rather than executable-level malware modifications or adaptive
sandbox-evasion strategies. Third, the DDQN component adapts only during offline training;
the deployed MLP is a fixed model that cannot adapt at runtime to new ransomware
behaviors, unseen families, or shifting attack patterns, thereby limiting responsiveness
to concept drift in real-world deployment~\cite{ref5}. Future work will address these
limitations through held-out family evaluation, external dataset validation, problem-space
adversarial testing, and runtime adaptation mechanisms, building on the principled and
trustworthy baseline established by the proposed framework.

Overall, the findings suggest that DDQN-guided adaptive weighting provides a credible and
practically efficient enhancement for behavioral ransomware detection while maintaining
lightweight deployment characteristics suitable for real-world security environments.

\section{Conclusion and Future Work}

The integration of three analytical components --- adaptive training-time weighting,
post-hoc explanation consistency, and adversarial robustness evaluation --- into a single
lightweight deployment pipeline constitutes the framework's most notable contribution to
the field of behavioral ransomware detection research. Rather than employing the DDQN
agent as a runtime detector, the proposed framework confines its application to the
training phase only. During this phase, it adaptively assigns sample weights based on the
batch-level loss and confidence metrics, whereas the final deployed detector is a
lightweight MLP. Under 5-fold stratified cross-validation, the DDQN--MLP achieves an
accuracy of $99.30\%$, an F1-score of $0.9930$, and an ROC-AUC of $0.9991$. This
represents a modest yet consistent improvement over the plain MLP, class-weighted MLP,
focal-loss MLP, classical models, and alternative RL variants, thereby affirming that
adaptive sample weighting enhances behavioral ransomware classification under heterogeneous
training conditions.

SHAP and LIME explanations confirmed that decisions were grounded in behaviorally
significant indicators, whereas SHAP--gradient alignment demonstrated broad consistency
between attribution patterns and classifier sensitivity. Adversarial training significantly
enhanced robustness against FGSM, BIM, JSMA, and PGD attacks across all tested magnitudes
($\varepsilon \in \{0.05, 0.10, 0.15, 0.20\}$), achieving an accuracy exceeding $98.6\%$
while fully maintaining clean-data performance. This confirms that trustworthiness and
deployment efficiency are not mutually exclusive in the proposed framework.

The findings should be interpreted within their evidential scope: the evaluation is
conducted in-distribution, adversarial experiments are limited to the feature space, and
generalization to unseen families, future variants, and external datasets has yet to be
established. Future research will address these limitations through held-out-family and
time-based evaluations, validation with external datasets, problem-space adversarial
testing, and runtime adaptation for continual deployment. Overall, the results position
DDQN-guided adaptive weighting as a principled and effective training-time strategy for
behavioral ransomware detection, while emphasizing that broader deployment claims require
further validation to be substantiated.

% Use the style of numbering in square brackets.
% If nothing is used, default style will be taken.
%\begin{enumerate}[a)]
%\item 
%\item 
%\item 
%\end{enumerate}  

%\begin{itemize}
%\item 
%\item 
%\item 
%\end{itemize}  

%\begin{description}
%\item[]
%\item[] 
%\item[] 
%\end{description}  

% Uncomment and use as the case may be
%\begin{theorem} 
%\end{theorem}

% Uncomment and use as the case may be
%\begin{lemma} 
%\end{lemma}


%\section{}\label{}

\printcredits

\bibliographystyle{cas-model2-names}

\bibliography{references}



\end{document}